\documentclass[a4paper,11pt]{article}
\usepackage[bottom=4.5cm,top=3.8cm,left=2.4cm,right=2.4cm]{geometry}
\usepackage{graphicx}
\usepackage{hyperref}

\usepackage{multirow}
\usepackage{tabularx}
\usepackage{booktabs}
\usepackage{url}
\newcolumntype{P}[1]{>{\centering\let\newline\\\arraybackslash\hspace{0pt}}p{#1}}

\usepackage{mathtools}
\usepackage{amsthm}
\usepackage{amsmath}
\usepackage{amssymb}
\usepackage{chngcntr}

\usepackage[shortlabels]{enumitem}

\usepackage{algorithmicx}
\usepackage{algpseudocode}
\usepackage{algorithm}

\newcommand{\floor}[1]{\lfloor{#1}\rfloor}

\newcommand{\dist}{{\mathop{\rm dist}}}

\newtheorem{theorem}{Theorem}[section]
\newtheorem{lemma}{Lemma}[section]
\newtheorem{property}{Property}[section]
\newtheorem{observation}{Observation}[section]
\newtheorem{assumption}{Assumption}[section]

\begin{document}
\begin{center}
{\Large Exact Set Packing in Multimodal Transportation with Ridesharing System for First/Last Mile}
\vskip 0.2in
Qian-Ping Gu$^1$, Jiajian Leo Liang$^1$

$^1$School of Computing Science, Simon Fraser University, Canada\\
qgu@sfu.ca, leo\_liang@sfu.ca
\end{center}

\begin{abstract}
We propose a centralized transportation system that integrates public transit with ridesharing to provide  multimodal transportation. At each time interval, the system receives a set of personal drivers, designated drivers, and public transit riders. It then assigns all riders to drivers, ensuring that pick-ups and drop-offs occur at designated transit stations. This effectively replaces first-mile/last-mile (FM/LM) segments with a ridesharing alternative, reducing overall commuting time.
We study two optimization problems: (1) minimizing the total travel distances of drivers and (2) minimizing the number of designated drivers required to serve all riders. We show the optimization problems are NP-hard and give hypergraph-based integer linear programming exact algorithm and approximation algorithms.
To enhance computational efficiency, we introduce a clustering heuristic that utilizes both spatial and temporal aspects of the input data to accelerate rider-to-driver assignments. Finally, we conduct an extensive computational study using real-world datasets and surveys from Chicago to evaluate our model and algorithms at a city-wide scale.
\end{abstract}


\section{Introduction} \label{sec-intro}
The transportation sector is responsible for about 25\% of global greenhouse gas 
emissions~\cite{Javed-PESCC19,IPCC22}. As urban and suburban areas expand, the number of commuters increases, 
leading to greater traffic congestion and environmental impact.
The inconvenience and inflexibility of the first mile (\textbf{FM}, commute from one's origin/home to the nearest transit stop or major transportation hub) and 
last mile (\textbf{LM}, commutes from a transit stop or major transportation hub to one's destination/home)  transportation, compared to personal vehicles, causes many people to choose to use personal vehicles for 
commutes~\cite{Burstlein-TRAPP21,Chen-TRBM18,Wang-TS14}.

Ridesharing and carpooling have been considered effective in reducing traffic congestion and extensively studied. 
Recent literature reviews on shared mobility can be found in~\cite{Mourad-TRBM19,Tafreshian-SS20,Wang-TRBM19}. 
An important research direction is to use ridesharing as a \textit{feeder} for public transit to improve the 
FM/LM efficiency. Several approaches have been proposed in this direction. These approaches include using 
autonomous vehicles to improve bus routes and allow flexible arrangement of transit vehicles by utilizing
reallocation and on-demand~\cite{Carrese-sus23,Fielbaum-IJITSR20,Huang-CEUS22,Pinto-TRCET20,Shen-TRAPP18}.
A recent approach to designing multimodal transportation systems includes ridesharing as an intrinsic part of the transit network~\cite{Fielbaum-TRCET24,Fielbaum-PT24,Fielbaum-TRAPP24}. 
Studies in~\cite{Gu-COR24,Ma-EEEIC17,Stiglic-COR18} use a graph matching approach to introduce ridesharing to replace FM/LM transit. Especially, the rider-driver assignment problem is modeled as a graph matching problem in \cite{Gu-COR24}.

In this paper, we propose a centralized multimodal transport system (MTRS) that aims to improve the FM/LM transportation to reduce the commuting time and traffic congestion. The system integrates the public transit with ridesharing using \textit{fixed transit stations} which are predefined intermediate stops, such as park-and-ride locations and major transportation hubs for pick-ups and drop-offs, where passengers can transfer between ridesharing services and public transit. Ridesharing providers consist of personal and designated drivers.

Personal drivers using their own vehicles for work commute can provide ridesharing service while commuting 
to/from work. This model offers a more convenient and efficient alternative to traditional FM/LM transit, and
multiple riders sharing a single vehicle can lower travel costs and improve access to public transit. As a result, 
transit ridership may increase and vehicles on the road may decrease, supporting recommendations 
from~\cite{Feigon-TRB16,Zhang-IJERPH18} for a more sustainable transportation system, particularly for the 
FM/LM transport through shared mobility services~\cite{MENG-TR20}. Designated drivers include mobility-on-demand (MoD) service drivers and taxi drivers.
MoD services, such as Uber, Lyft and DiDi, have become popular around the globe for their convenience. As ridesharing and ride-hailing services continue to grow, several studies suggest that integrating MoD with public transportation could enhance urban mobility and increase public transit ridership~\cite{Alonso-Gonzalez-TRR18,Hall-JUE18,Kumar-TRCET21,Narayan-TRCET20}. By participating in the transportation system, personal drivers can offset their driving costs, while designated drivers can increase their earnings.

The concept of large mobility transportation hubs, which facilitate multimodal travel, has gained popularity in 
research and policy making as shown in some recent studies (e.g.,~\cite{Blad-CSTP22,Rongen-JTG22}).
Fixed drop-off transit stations enhance route visibility and improve schedule predictability for personal drivers. Additionally, park-and-ride facilities and large transportation hubs with parking allow 
personal drivers to switch to public transit for part of their journey. This hybrid approach can provide 
a faster commute, particularly when using subways or trains to bypass traffic congestion. If the integrated 
transportation system proves to be cost-effective and reliable, many personal drivers may transition away 
from private vehicle use in favor of this multimodal solution. 

Our system aims at minimizing the total travel distance of drivers and the number of drivers to serve all riders while improving FM/LM efficiency, leading to a more efficient and sustainable urban mobility network. Specifically, we focus on two optimization problems:
\begin{enumerate}
\item[(1)] Minimizing total travel distance (\textit{MTRS-minDist}): Assign all riders to drivers such that each rider's original commuting time is reduced by a predefined value and the sum of detour distances of personal drivers and 
travel distances of designated drivers is minimized.
\item[(2)] Minimizing the number of designated drivers (\textit{MTRS-minNum}): Assign all riders to drivers such that each rider's original commuting time is reduced by a predefined value and the number of designated drivers used is minimized. 
\end{enumerate}
To solve these optimization problems, we adopt a similar model to that in~\cite{Gu-COR24,Stiglic-COR18}, 
providing exact algorithm and approximation algorithms, which are similar to finding a set packing that covers all riders. Our main contributions are summarized as follows:
\begin{enumerate}\setlength\itemsep{0em}
\item We give a hypergraph-based integer linear program (ILP) exact algorithm approach for solving the 
MTRS-minDist and MTRS-minNum problems, and we show that both optimization problems are NP-hard.

\item We give a $\frac{\lambda^2\mu + \lambda}{\lambda + 1}$-approximation algorithm for MTRS-minDist, where $\lambda$ is the maximum vehicle capacity and $\mu$ is the ratio of the longest to shortest incurred-travel-distance ridesharing routes.
For MTRS-minNum, we give a $\frac{\lambda+2}{2}$-approximation algorithm and a local search heuristic.

\item We introduce a clustering heuristic that reduces the computational time needed to compute the hypergraph while preserving solution quality.

\item We conduct an extensive computational study to evaluate the system and optimization algorithms using
real-life data from Chicago City.
The total travel distance of drivers and number of drivers obtained by our approximation algorithms are close to the optimal solutions while reducing commute time more than the predefined value of 30\%.
\end{enumerate}

For the remainder of this paper, Section~\ref{sec-preliminary} describes our centralized system and defines the MTRS-minDist and MTRS-minNum problems. Then we present the ILP exact algorithm and the NP-hardness results.
In Section~\ref{sec-mtrs-algorithms}, we present approximation algorithms for MTRS-minDist and MTRS-minNum, 
followed by the clustering heuristic in Section~\ref{sec-clustering}. We discuss our numerical experiments and results in Section~\ref{sec-experiment}. Finally, Section~\ref{sec-conclusion} concludes the paper.

\section{Problem definition and preliminaries} \label{sec-preliminary}
For each time interval, the multimodal transportation with ridesharing system (MTRS) receives a set 
$R=\{r_1,\ldots,r_{|R|}\}$ of riders who request ridesharing service and a set $D=\{\eta_1,\ldots,\eta_{|D|}\}$ 
of drivers who offer ridesharing service. A public transit network with a set $TS$ of \emph{transit stations} 
(\emph{station} for brevity) and a timetable $T$ is given.  
Riders of $R$ can be picked up at their origins and dropped off at the stations or picked up at the stations and dropped off at their destinations, by drivers in $D$. A \emph{public transit route} for rider $r_j\in R$ is a travel plan using only public transportation, whereas a \emph{ridesharing route} for rider $r_j$ is a travel plan using a combination of public transportation and ridesharing to reach $r_j$'s destination. An edge-weighted directed graph $G(V,E,w)$ (road network) is given, which represents locations reachable by both private and public transportation.

The set $D=\Gamma\cup \Delta$ of drivers consists of a set $\Gamma$ of personal drivers who drive their vehicles for work commute and a set $\Delta$ of designated drivers who are employed/self-employed to offer ridesharing service with $\Gamma\cap \Delta=\emptyset$. Personal drivers are on the road independent of ridesharing services, whereas designated drivers are on the road only if necessary (e.g., Uber and Taxi drivers).
Each driver $\eta_i \in D$ has travel requirements, including origin $o_i$, destination $d_i$, earliest departure time $\alpha_i$ from $o_i$, latest arrival time $\beta_i$ at $d_i$, the vehicle capacity $\lambda_i$ available for riders, and maximum detour time limit $z_i$ to serve riders. 
Each rider $r_j \in R$ also has travel requirements with $o_j,d_j,\alpha_j,\beta_j$ as defined in the travel requirements of drivers, and minimum time saved $\theta_j$ for $r_j$ to use a ridesharing route (compared to the fastest public transit route). A ridesharing route satisfying a rider $r_j$'s requirements is called a \emph{feasible route} for $r_j$.
Two match types are considered in this paper:

\begin{itemize} \setlength\itemsep{0em}
\item Match type \textbf{FM} (first mile): a driver may make stops to pick up multiple riders but makes only one stop to drop off all riders. The \emph{pick-up locations} are the riders' origins, and the \emph{drop-off location} is a station 
in $TS$.
\item Match type \textbf{LM} (last mile): a driver makes only one stop to pick up riders and may make multiple stops to drop off all riders. The \emph{pick-up location} is a station in $TS$, and the \emph{drop-off locations} are the riders' destination.
\end{itemize}
Drivers and riders specify which match type to participate in as inputs to the system.
A driver (rider) who specifies FM match type is called a FM-driver (FM-rider), and similarly for LM-driver and LM-rider.
A driver $\eta_i$ \emph{can serve} rider $r_j$ if a feasible route exists for $r_j$ offered by $\eta_i$, $\eta_i$ and $r_j$ have the same match type (FM or LM), and the requirements of $\eta_i$ are also satisfied for serving $r_j$. A driver $\eta_i$ \emph{can serve} a set $R_i \subseteq R$ of riders means that $\eta_i$ and all riders of $R_i$ have the same match type, there is a feasible route offered by $\eta_i$ that satisfies the requirements of $\eta_i$ and all riders in $R_i$ collectively.

For two locations $o$ and $d$ in road network $G$, let $FP(o,d)$ be the fastest path from $o$ to $d$.
Let $R_i$ be the set of riders served by driver $\eta_i$, and let $\bar{P}(\eta_i, R_i)$ be the ridesharing route in $G$ that $\eta_i$ uses to serve $R_i$, which starts at $o_i$ and ends at $d_i$.
For a path $P$ in $G$, let $\dist(P)$ be the distance of $P$.
The \emph{incurred travel distance} of a personal driver $\eta_i \in \Gamma$ is the detour distance $\eta_i$ needs to travel to serve $R_i$, which is $TD(\eta_i, R_i)=\dist(\bar{P}(\eta_i, R_i)) - \dist(FP(o_i, d_i))$.
The \emph{incurred travel distance} of a designated driver $\eta_i \in \Delta$ is $TD(\eta_i, R_i) = \dist(\bar{P}(\eta_i, R_i))$.
The travel distance $TD(\eta_i, R_i)$ is represented by an integer, in meter.
The above definitions and notation are summarized in Table~\ref{table-problem-notation}.
\begin{table}[!t]
\small
%\footnotesize
%\setlength\tabcolsep{4pt}
\centering
   \begin{tabular}{c | l}
   	\hline
   	\textbf{Notation} & \textbf{Definition}                                          \\ \hline
   	$o_i$             & Origin (start location) of driver $\eta_i$ or rider $r_i$ (a vertex in road network $G$). \\
   	$d_i$             & Destination of driver $\eta_i$ or rider $r_i$  (a vertex in $G$).  \\
   	$\alpha_i$        & Earliest departure time of driver $\eta_i$ or rider $r_i$.          \\
   	$\beta_i$         & Latest arrival time of driver $\eta_i$ or rider $r_i$.              \\
	$\lambda_i$       & Vehicle capacity of $\eta_i$, number of seats available for riders.  \\
   	$z_i$             & Maximum detour time $\eta_i$ is willing to spend for offering ridesharing.    \\
    $\theta_i$        & Acceptance threshold ($0 < \theta_i \leq 1$) for a ridesharing route offered to rider $r_i$.  \\
    $\lambda$         & $\lambda = \max_{\eta_i \in D} \lambda_i$ is the largest vehicle capacity of all drivers.	\\
    $FP(o,d)$  	   & The fastest path from location $o$ to location $d$ in $G$. 		\\
    $\dist(P(u,v))$	   & The distance of path $P(u,v)$ from $u$ to $v$ in $G$. \\
    $\bar{P}(\eta_i, R_i)$ & The ridesharing route in $G$ that $\eta_i$ uses to serve all riders of $R_i$. \\
    $TD(\eta_i, R_i)$  & The travel distance of a driver $\eta_i$ to serve all riders of $R_i$, represented by an integer. \\
    \hline
   \end{tabular}
\caption{Basic definitions, notation, and the parameters of a driver $\eta_i$ or a rider $r_i$.}
\label{table-problem-notation}
\end{table}

We study two optimization problems. For feasibility and solvability, we assume the following.
\begin{assumption}\label{assumption-drivers}
Every rider of $R$ can be assigned a feasible route offered by a distinct driver of $\Delta$.
\end{assumption}
The first optimization problem (\textbf{MTRS-minDist}): 
\begin{itemize}
\item Given a set $D$ of drivers and a set $R$ of riders, the system assigns a feasible route for every rider of $R$ offered by some driver of $D$ such that each driver belongs to at most one assigned feasible route and each rider receives exactly one feasible route, and the total travel distance of drivers is minimized.
\end{itemize}
The second optimization problem (\textbf{MTRS-minNum}) focuses on minimizing the number of vehicles on the road. Since personal drivers are on the road regardless, we assume a maximal number of riders has been assigned to personal drivers in $\Gamma$ by some algorithm, exact or approximate (assigning the maximum number is NP-hard~\cite{Bei-AAAI18,Gu-COR24}).
MTRS-minNum is to minimize the number of drivers in $\Delta$ needed to serve all remaining riders 
$R' \subseteq R$.
More formally:
\begin{itemize}
\item Given a set $D = \Delta$ of designated drivers and a set $R'$ of riders, the system assigns a feasible route for every rider of $R'$ offered by some driver of a set $D' \subseteq D$ such that each driver belongs to at most one assigned feasible route, each rider receives exactly one feasible route, and $|D'|$ is minimized.
\end{itemize}

\subsection{Exact algorithm} \label{subsec-exact}
For a driver $\eta_i \in D$ and a group of passengers $R_i \subseteq R$, $(\eta_i,R_i)$ is a \emph{feasible match} if $\eta_i$ can serve all riders of $R_i$ by using a feasible route, which is a path in $G$ to serve all riders of $R_i$. Each feasible match $(\eta_i,R_i)$ is associated with an integer weight $w(\eta_i,R_i)$, depending on the optimization goal.
For MTRS-minDist, $w(\eta_i,R_i) = TD(\eta_i, R_i)$, as defined previously. For MTRS-minNum, $w(\eta_i,R_i) = 1$ (note that $\eta_i \in \Delta$).

The exact algorithm for both problems is as follows:
First, we compute all feasible matches for each driver $\eta_i \in D$ as shown in~\cite{Gu-COR24,Stiglic-COR18}.
Then, we construct an edge-weighted hypergraph $H(U \cup V,E,w)$, where $U(H) = D$ and $V(H) = R$ are the vertices.
For each $\eta_i \in D$ and for every subset $R_i$ of $R$, create a hyperedge $e=\{\eta_i\} \cup R_i$ in $E(H)$ if $(\eta_i, R_i)$ is a feasible match.
Each edge $e=\{\eta_i\} \cup R_i \in E(H)$ has weight $w(e)=w(\eta_i,R_i)$.
For a vertex $u \in U(H) \cup V(H)$, define $E_H(u) = \{e \in E(H) \mid u \in e\}$ to be the set of edges in $E(H)$ incident to $u$ (containing $u$); we also use $E(u)$ to denote $E_H(u)$ if $H$ is clear from the context.
It was noted in~\cite{Gu-ISAAC21} that Observation~\ref{obs-1} holds for $H(V\cup U,E,w)$.
\begin{observation}
For any edge $\{\eta_i\}\cup R_i \in E(H)$ with $|R_i|\geq 2$, the edge $\{\eta_i\}\cup R'_i$ exists in $E(H)$ for every non-empty subset $R'_i$ of $R_i$.
\label{obs-1}
\end{observation}

To solve the MTRS-minDist and MTRS-minNum problems, we give an integer linear program (ILP) formulation:
\begin{alignat}{4}
 & \text{minimize }  &    &\sum_{e \in E(H)} w(e) \cdot x_{e}  & \qquad \tag{\romannumeral1} \label{formulation-minimization} \\
 & \text{subject to } & \qquad &\sum_{e \in E(\eta_i)} x_{e} \leq 1,  & & \forall \text{ } \eta_i \in V(H) \tag{\romannumeral2}\label{constraint-driver}\\
 &                           & \qquad  &\sum_{e \in E(r_j)} x_{e} = 1, & &\forall \text{ } r_j \in U(H) = R \tag{\romannumeral3} \label{constraint-rider} \\
 &                            &        &x_{e} \in \{0,1\}, & &\forall \text{ } e \in E(H) \tag{\romannumeral4} \label{constraint-binary}
\end{alignat}
For the binary variable $x_e$ corresponding to $e=\{\eta_i\}\cup R_i$, if $x_e = 1$ then all riders of $R_i$ are assigned to $\eta_i$, indicating all riders of $R_i$ are delivered by $\eta_i$.
The objective function~\eqref{formulation-minimization} is to minimize the total incurred travel distance of the drivers (for MTRS-minDist) or to minimize the number of designated drivers (for MTRS-minNum) needed to serve all riders.
Constraint~\eqref{constraint-driver} %in the ILP formulation 
guarantees that each driver serves at most one feasible set of riders, and each rider is served by exactly one driver by constraint~\eqref{constraint-rider}.
A feasible solution to MTRS-minDist and MTRS-minNum is a subset $E' \subseteq E(H)$ with $R(E') = R$.

\subsection{NP-hard results}
Let $\lambda=\max_{\eta_i\in D} \lambda_i$. If $\lambda = 1$, $H$ is a bipartite graph, and both MTRS-minDist and MTRS-minNum can be solved in polynomial time by finding a minimum weight maximum matching on $H$~\cite{Ahuja-NF93}.
We assume $\lambda > 1$ hereafter. Then our problems are variants of the exact set packing problem. If every edge of $H$ has uniform weight, our problems are similar to exact set packing, except edges of $H$ satisfy Observation~\ref{obs-1} and only all riders need to be covered.
If $\lambda > 1$, both MTRS-minDist and MTRS-minNum are NP-hard; and we prove this by reducing from the 3-dimensional perfect matching (3DM) problem.
An instance of 3DM consists of three disjoint finite sets $A$, $B$ and $C$, and a collection $\mathcal{F} \subseteq A \times B \times C$.
That is, $\mathcal{F}$ is a collection of triplets $(a,b,c)$, where $a \in A, b \in B$ and $c \in C$.
A 3-dimensional matching is a subset $\mathcal{M} \subseteq \mathcal{F}$ such that all triplets in $\mathcal{M}$ are pairwise disjoint.
The decision problem of 3DM is that given $(A, B, C, \mathcal{F})$ and an integer $q$, decide whether there exists a matching $\mathcal{M} \subseteq \mathcal{F}$ with $|\mathcal{M}| \geq q$, which is NP-complete.
Given an instance $(A,B,C,\mathcal{F})$ of 3DM with $|A| = 2q$ and $|B| = |C| = q$, we construct a hypergraph  $H(U\cup V,E,w)$ for an instance of the MTRS-minDist problem as follows:
\begin{itemize}
\setlength\itemsep{0em}
\item Create a set of drivers $D(H) = A$ with capacity $\lambda_i=2$ for every driver $\eta_i \in D(H)$ and a set of riders $R(H) = B \cup C$.

\item For each $f \in \mathcal{F}$, create a hyperedge $e(f)$ in $E(H)$ containing elements $\{a,b,c\}$, where $a $ represents a driver in $D(H)$, and $b\in B$ and $c\in C$ represent two different riders in $R(H)$.
Further, create and add edges $e'(f) = \{a, b\}$ and $e''(f) = \{a, c\}$ to $E(H)$ so that Observation~\ref{obs-1} is maintained.
Assign weights $w(e(f)) = w(e'(f)) = w(e''(f)) = 2$.
\item For every pair $a \in A$ and $b \in B$, add an edge $e=\{a, b\}$ with $w(e) = \omega \geq 2$ to $E(H)$ if $e$ is not already in $E(H)$; and for every pair $a \in A$ and $c \in C$, add an edge $e=\{a, c\}$ with $w(e) = \omega$ to $E(H)$ if $e$ is not already in $E(H)$ so that Assumption~\ref{assumption-drivers} is satisfied.
\end{itemize}
Note that the above construction also satisfies Assumption~\ref{assumption-designated-drivers-strong}.

\begin{lemma} \label{lemma-minDist-nphard}
An instance $(A,B,C,\mathcal{F})$ of the maximum 3-dimensional matching problem has a solution $\mathcal{M}$ of cardinality $q$ if and only if the objective function value of ILP~\eqref{formulation-minimization}-\eqref{constraint-binary} for the hypergraph $H(U\cup V,E,w)$ is $2q$.
\end{lemma}

\begin{proof}
Assume that $(A,B,C,\mathcal{F})$ has a solution $\mathcal{M} = \{f_1, f_2,\ldots, f_q\}$. For each $f_i$ ($1 \leq i \leq q$), set the corresponding binary variable $x_{e(f_i)} = 1$ in ILP~\eqref{formulation-minimization}-\eqref{constraint-binary}.
Since $f_i \cap f_j = \emptyset$ for $1 \leq i \neq j \leq q$, constraint~\eqref{constraint-driver} of the ILP is satisfied for every $\eta_i = a \in f_i$.
Further, $|\mathcal{M}| = q$ and each $f_i$ contains exactly one element from $B$ and one element from $C$, implying constraint~\eqref{constraint-rider} of the ILP.
Since each edge $e(f_i)$ corresponding to $f_i\in \mathcal{M}$ has weight $w(e(f_i))=2$, the objective function value of ILP~\eqref{formulation-minimization}-\eqref{constraint-binary} is $2q$.

Assume that the objective function value of ILP~\eqref{formulation-minimization}-\eqref{constraint-binary} is $2q$.
Let $X=\{e(f) \in E(H) \mid x_{e(f)} = 1\}$, where $x_{e(f)}$'s are the binary variables of the ILP.
For every edge $e(f) \in X$, add the corresponding set $f \in \mathcal{F}$ to $\mathcal{M}$.
From constraints~\eqref{constraint-driver} and \eqref{constraint-rider} of the ILP, $X$ is pairwise vertex-disjoint and $|X| \leq |D(H)|$.
For the objective function value to be $2q$, $|X| = q$ since each edge in $E(H)$ has weight at least 2.
Recall that every $e(f) \in X$ contains at most two different riders and all riders of $R(H)$ must be served by constraint \eqref{constraint-rider}.
To serve all $2q$ riders with $q$ edges, each edge in $X$ must contain exactly 2 distinct riders.
Hence, $\mathcal{M}$ is a valid solution for $(A,B,C,\mathcal{F})$ with $|\mathcal{M}| = |X| = q$.

The size of $H(V,E)$ is polynomial in $q$. It takes a polynomial time to convert a solution of $H(V,E)$ to a solution of the 3DM instance $(A,B,C,\mathcal{F})$ and vice versa.
\end{proof}

\begin{theorem} \label{theorem-minDist-nphard}
The MTRS-minDist problem is NP-hard.
\end{theorem}

\begin{proof}
The ILP~\eqref{formulation-minimization}-\eqref{constraint-binary} always finds a feasible solution to the MTRS-minDist problem by constraints~\eqref{constraint-driver} and \eqref{constraint-rider} of the ILP, and with the objective function value, the ILP~\eqref{formulation-minimization}-\eqref{constraint-binary} always finds an optimal solution.
It is obvious that an optimal solution to the MTRS-minDist problem is an optimal solution to the ILP~\eqref{formulation-minimization}-\eqref{constraint-binary} by definition.
Lemma~\ref{lemma-minDist-nphard} implies that it is NP-hard to find an optimal solution to the ILP~\eqref{formulation-minimization}-\eqref{constraint-binary} unless P=NP. Hence, the MTRS-minDist problem is NP-hard.
\end{proof}

The NP-hardness of MTRS-minNum problem can be proved in a similar way.
Construct a hypergraph $H(U\cup V,E,w)$ of an instance of the MTRS-minNum problem as described above for MTRS-minDist, expect the weight of each edge in $E(H)$ is 1.
With a very similar analysis of Lemma~\ref{lemma-minDist-nphard}, we have the following lemma.

\begin{lemma} \label{lemma-minNum-nphard}
An instance $(A,B,C,\mathcal{F})$ of the maximum 3-dimensional matching problem has a solution $\mathcal{M}$ of cardinality $q$ if and only if the objective function value of ILP~\eqref{formulation-minimization}-\eqref{constraint-binary} for the hypergraph $H(U\cup V,E,w)$ is $q$.
\end{lemma}

Then from Lemma~\ref{lemma-minNum-nphard}, Theorem~\ref{theorem-minNum-nphard} follows by a similar analysis as in Theorem~\ref{theorem-minDist-nphard}.

\begin{theorem} \label{theorem-minNum-nphard}
The MTRS-minNum problem is NP-hard.
\end{theorem}

Hypergraph $H(U \cup V,E,w)$ and notations related to $H$ below will be used in the rest of the paper. 
Let $D(H) = U(H) \cap D$, $\Gamma(H) = U(H) \cap \Gamma$ and $\Delta(H) = U(H) \cap \Delta$.
For any edge $e \in E(H)$, let $N_H(e)$ be the set of edges in $E(H)$ incident to $e$ ($N(e)$ when $H$ is clear from the context).
For any edge $e=\{\eta_i\} \cup R_i$ in $H$, let $D(e) = \eta_i$ and $R(e) = R_i$.
For any subset $E' \subseteq E(H)$, let $R(E') = \bigcup_{e \in E'} R(e)$, $D(E') = \bigcup_{e \in E'} D(e)$ and $w(E') = \sum_{e \in E'} w(e)$.
Note that $\Gamma(E') = D(E')\cap \Gamma$, and likewise, $\Delta(E') = D(E')\cap \Delta$.

\section{Algorithms for MTRS-minDist and MTRS-minNum} \label{sec-mtrs-algorithms}
Based on Assumption~\ref{assumption-designated-drivers-strong}, we give algorithms for MTRS-minDist and MTRS-minNum that achieve approximation ratios better than trivial upper bounds. Both problems remain NP-hard even if Assumption~\ref{assumption-designated-drivers-strong} is satisfied, as shown in the NP-hardness proof above.

\begin{assumption}\label{assumption-designated-drivers-strong}
$|\Delta|=|R|$ and for every $\eta_i \in \Delta$ and every $r_j\in R$, there is a feasible route for $r_j$ offered by $\eta_i$.
\end{assumption}

It is worth to mention that our algorithms for MTRS-minDist and MTRS-minNum are modified from algorithms for the maximum (weighted) k-set packing problem~\cite{Chandra-JoA01,Furer-ISCO14}. Those algorithms do not apply to our optimization problems because not all riders are guaranteed to be served (such as algorithms in \cite{Chandra-JoA01}) or weight is not considered (e.g.,~\cite{Furer-ISCO14}).

\subsection{Algorithm GreedyMinDist}
Let $H(U \cup V,E,w)$ be the hypergraph constructed for an instance of the MTRS-minDist problem.
Algorithm GreedyMinDist computes two solutions $M_1$ and $M_2$ separately, and then selects 
either $M_1$ or $M_2$ as the solution $M$ with $w(M)=\min\{w(M_1),w(M_2)\}$.

\begin{itemize}[align=left]
\item[(1)] $M_1$ is computed as follows with the original $H(U \cup V,E,w)$.\\
Initially, $M_1=\emptyset$. In each iteration, select an edge $e$ in $E(H)$ with minimum $w(e)$, $M_1=M_1\cup \{e\}$, remove
all vertices of $e$ from $U(H)\cup V(H)$ and all edges of $\cup_{u \in e} E(u)$ from $E(H)$.
Continue to the next iteration until $R(M_1) = R$.
\end{itemize}

\begin{itemize}
\item[(2)] $M_2$ is computed as follows with the original $H(U \cup V,E,w)$.\\
For each edge $e \in E(H)$, compute ratio $w'(e) = w(e) / |R(e)|$.
Initially, $M_2=\emptyset$.
In each iteration, selects an edge $e$ in $E(H)$ with minimum $w'(e)$, $M_2=M_2\cup \{e\}$, and removes 
all vertices of $e$ from $U(H) \cup V(H)$ and all edges of $\cup_{u \in e} E(u)$ from $E(H)$.
Continue to the next iteration until $R(M_2) = R$.
\end{itemize}
Algorithm GreedyMinDist has at most $|R|$ iterations for each solution computed.
Let $M$ be the solution computed by GreedyMinDist and $M^*$ be an optimal solution to the MTRS-minDist problem.
A driver $\eta_u$ or a rider $r_v$ is said to be \emph{covered} by some solution if an edge $e$ containing 
$\eta_u$ or $r_v$ is in the solution; otherwise, \emph{uncovered} with respect to (w.r.t.) the solution.
First, we show GreedyMinDist always finds a feasible solution.

\begin{lemma} \label{lemma-Greedy-feasible}
Algorithm GreedyMinDist always computes a feasible solution $M$ to the MTRS-minDist problem.
\end{lemma}

\begin{proof}
Let $M=\{e_1,e_2\ldots,e_{|M|}\}$, where $e_i$ is the $i^{th}$ edge added to $M$.
After the edge $e_i$ is added to $M$ at iteration $i$, all the edges in $H$ incident to $e_i$ are removed from $H$, implying all edges of $M$ are pairwise vertex-disjoint.
In addition, exactly one driver (which is $D(e_i)$) is removed from $U(H)$ at the end of each iteration $i$.
From this and Assumption~\ref{assumption-designated-drivers-strong}, for each uncovered rider $r_v \in R$ w.r.t. current solution $M$, there must exist a distinct driver $\eta_u$ that can serve $r_v$, namely, $\{\eta_u, r_v\}$ remains in $H$ before $r_v$ is covered.
Hence, every rider will be covered when the algorithm terminates.
\end{proof}

Next, we show the approximation ratio of Algorithm GreedyMinDist.
Let $\mu = \frac{max_{e\in E(H)} w(e)}{min_{e\in E(H)} w(e)}$.
Recall that $w(e) = TD(\eta_i, R_i)$ is an integer, and in general, a shorter-distance path has a shorter travel time and vice verse, implying $w(e)$ is almost always positive.
For analysis purpose, we assume that the more locations a driver needs to visit, the longer the distance and time are required, which implies that $w(e) \geq 1$ for all edges $e$ in $H$ and the the following property.

\begin{property} \label{property-subset-weight}
For any pair of edges $e$ and $e'$ in $E(H)$ such that $D(e) = D(e')$ and $R(e') \subseteq R(e)$, $w(e') \leq w(e)$.
\end{property}

\begin{lemma} \label{lemma-extreme-solution}
A solution $M'$ can be constructed from $M_1$ such that $|R(e')|=1$ for every $e'\in M'$, $w(M') \geq w(M_1)$ and $M'$ is a solution computed by Step (1) of Algorithm GreedyMinDist with restriction to selecting edges containing exactly one rider.
\end{lemma}

\begin{proof}
For every edge $e_i \in M_1$ with $|R(e_i)|= 1$, add $e_i$ to $M'$.
For every edge $e_i \in M_1$ with $|R(e_i)| > 1$, split $e_i$ into a set $M'(e_i) = \{e'_{i_1},e'_{i_2},\ldots,e'_{i_z}\}$ of edges (where $z =|R(e_i)|$) as follows.
For each $e'_{i_j}$, $|R(e'_{i_j})| = 1$ for $1 \leq j \leq z$.
$D(e'_{i_j}) \neq D(e'_{i_l})$ for $1 \leq j \neq l \leq z$, $D(e'_{i_1}) = D(e_i)$, and $D(e'_{i_j}) \notin D(M_1) \cup D(M')$ for $2 \leq j \leq z$.
By Assumption~\ref{assumption-designated-drivers-strong}, edges of $M'(e_i)$ can be assigned such that $R(M'(e_i)) = R(e_i)$.
Add $M'(e_i)$ to $M'$.
From Property~\ref{property-subset-weight} and $e_i$ is selected instead of $e'_{i_1}$, $w(e'_{i_1}) = w(e_i)$.
This implies that $w(M'(e_i)) \geq w(e_i)$ and $w(M') \geq w(M_1)$.
Partition $M'$ into two sets $M'_1$ and $M'_2$, where $D(M'_1) = D(M)$ and $M'_2 = M' \setminus M'_1$.
The actual edges of $M'_2$ can be selected by following Step (1) of GreedyMinDist for edges containing exactly one rider.
The selection of $M'_2$ does not affect $M'_1$ and vice versa since the edges in $M'$ are pairwise vertex-disjoint.
Order all the edges of $M'=\{e'_1,\ldots, e'_{|M'|}\}$ such that $w(e'_j) \leq w(e'_l)$ for $1 \leq j < l \leq |M'|$.
Then, $M'$ is a solution computed by Step (1) of GreedyMinDist for selecting edges containing exactly one rider.
\end{proof}

By Lemma~\ref{lemma-extreme-solution}, we can assume $M_1$ computed by GreedyMinDist has property that $|R(e)|=1$ for every $e\in M_1$. Notice that $|D(M_1)|=|R|$ and $D(M_1)=\Delta$.
For the analysis purpose, we make every edge $e^*$ of $M^*$ contains exactly $\lambda$ riders, as follows.
For every edge $e^* \in M^*$ with $|R(e^*)| < \lambda$, add $\lambda - |R(e^*)|$ dummy riders to $e^*$ such that $w(e^*)$ remains unchanged, and add $\lambda - |R(e^*)|$ dummy edges $\{\eta_r\} \cup r$ to $M_1$ with $w(\eta_r, r) = 0$, where $\eta_r$ is the dummy driver uniquely associated with the dummy rider $r$.
After the construction, $w(M^*)$ and $w(M_1)$ remain unchanged, so the ratio $w(M_1)/w(M^*)$ remains unchanged. Further, both solutions remain feasible to the origin problem instance (by removing all dummy riders and drivers).
Notice that it remains true that $w(e^*) \geq min_{e\in E(H)} w(e)$ and $w(e^*) \leq max_{e\in E(H)} w(e)$ for every $e^* \in M^*$ since $w(e^*)$ is unchanged; and $w(e) \leq max_{e\in E(H)} w(e)$ for every edge $e \in M_1$.

We first show that a subset $\hat{M} \subseteq M_1$ and a subset $\hat{M}^* \subseteq M^*$ with $|\hat{M}| = |\hat{M}^*|$ that can be paired with each other in Lemma~\ref{lemma-GreedyMinDist-weight}.
\begin{lemma} \label{lemma-GreedyMinDist-weight}
There is a non-empty subset $\hat{M} \subseteq M_1$ and a non-empty subset $\hat{M}^* \subseteq M^*$ such that $|\hat{M}| = |\hat{M}^*|$, $w(\hat{M}) \leq w(\hat{M}^*)$ and $|\hat{M}^*|\geq \lambda|M^*|/(\lambda + 1)$.
\end{lemma}

\begin{proof}
We describe how $\hat{M}$ and $\hat{M}^*$ can be constructed.
Notice that $\Delta \subseteq D(M_1)$ (due to dummy drivers), and this implies that $D(M^*) \subseteq D(M_1)$.
For each $e^* \in M^*$, there is exactly one edge $e\in M_1$ with $D(e) = D(e^*)$.
If $w(e) \leq w(e^*)$ then $e$ is \textit{paired with} $e^*$, denoted by $f(e) = e^*$.
Let $\hat{M} \subseteq M$ and $\hat{M}^* \subseteq M^*$ such that for every $e \in \hat{M}$, there is an edge $e^* \in \hat{M}^*$ with $f(e) = e^*$, $D(e)=D(e^*)$ and $w(e) \leq w(e^*)$.
The function $f: \hat{M} \rightarrow \hat{M}^*$ is bijective since for every edge $e \in \hat{M}$, there is exactly one edge $e^* \in \hat{M}^*$ with $D(e)=D(e^*)$, and vice versa.
For each edge $e^{*} \in M^*$ not paired, if $e^*$ contains a dummy rider $r$, then there exists exactly one dummy edge $e=\{\eta_r\} \cup r$ in $M_1 \setminus \hat{M}$ with $w(e) = 0$.
We pair this edge $e$ with $e^*$ (that is, $f(e) = e^*$), and include $e$ in $\hat{M}$ and $e^*$ $\hat{M}^*$.
Since $e$ and $e^*$ are uniquely paired with each other, $f: \hat{M} \rightarrow \hat{M}^*$ remains bijective.

For each edge $e^{**} \in M^*$ not paired, the edge $e\in M_1$ with $D(e)=D(e^{**})$ has weight $w(e) > w(e^*)$.
For every rider $r \in R(e^{**})$, there is an edge $e'_r = \{D(e^{**}), r\}$ in $H$ by Observation~\ref{obs-1} and $w(e'_r) \leq w(e^{**})$ by Property~\ref{property-subset-weight}.
Since edge $e$ has weight $w(e) > w(e^*)$, $e'_r$ does not cover any rider of $R(e^{**})$, $e'_r \notin M_1$ and every rider $r \in R(e^{**})$ is covered by an edge $e''_r \in M_1$. Before edge $e''_r$ is added to $M_1$, $e'_r$ is a valid selection for the algorithm ($e$ has not been added to $M_1$ yet), implying $w(e''_r) \leq w(e'_r) \leq w(e^{**})$.
Each edge $e^{**} \in M^*$ has not been paired is considered one by one in an arbitrary order, where each $e^{**}$ is considered w.r.t. an updated $\hat{M}$ and $\hat{M}^*$.
If $R(e^{**}) \setminus R(\hat{M}) \neq \emptyset$, then there is a rider $r \in R(e^{**}) \setminus R(\hat{M})$ and an edge $e''_r \in M_1 \setminus \hat{M}$ which covers $r$ with $w(e''_r) \leq w(e^{**})$ as shown above.
We pair $e''_r$ with $e^{**}$, denoted by $f(e''_r) = e^{**}$.
Since $e''_r \notin \hat{M}$ and $e^{**} \notin \hat{M}^*$, $e''_r$ and $e^{**}$ are uniquely paired with each other.
Then include $e''_r$ in $\hat{M}$ and $e^{**}$ in $\hat{M}^*$, and $f: \hat{M} \rightarrow \hat{M}^*$ remains bijective.
Further, $w(e) \leq w(e^*)$ and $w(e''_r) \leq w(e^{**})$ for every $e, e''_r\in \hat{M}$ and the corresponding $f(e)=e^*$ and $f(e''_r)=e^{**}$ in $\hat{M}^*$.

Hence, $|\hat{M}| = |\hat{M}^*|$ and $w(\hat{M}) \leq w(\hat{M}^*)$ by the above construction.
For each edge $e^* \in M^* \setminus \hat{M}^*$, $R(e^*) \subseteq R(\hat{M})$.
Since $|R(e^*)| = \lambda$ for every $e^* \in M^*$, $|\hat{M}^*| = |\hat{M}| \geq \lambda(|M^*| - |\hat{M}^*|)$, implying 
$|\hat{M}^*|(\lambda + 1) \geq \lambda|M^*|$.
\end{proof}

\begin{theorem} \label{theorem-minDist-Greedy-app}
Let $M$ be a solution found by Algorithm GreedyMinDist and $M^*$ be an optimal solution to the MTRS-minDist problem. Under Assumption~\ref{assumption-designated-drivers-strong} and $w(e) \geq 1$ for all edges $e$ in $H$, $\frac{w(M)}{w(M^*)} \leq \frac{\lambda^2\mu + \lambda}{\lambda + 1}$ for $\lambda \geq 2$.
\end{theorem}

\begin{proof}
Suppose $M = M_1$.
From Lemma~\ref{lemma-GreedyMinDist-weight}, $w(\hat{M}) \leq w(\hat{M}^*)$ and each rider of $R(M^*)\setminus R(\hat{M})$ is covered by an edge $e \in M\setminus \hat{M}$ with weight at most $max_{e\in E(H)} w(e)$.
Hence,
\begin{align} \label{eq-base-weight}
w(M) &\leq |R(M^*)\setminus R(\hat{M})| \max_{e\in M} w(e) + w(\hat{M}) \nonumber \\
&\leq |R(M^*)\setminus R(\hat{M})| \max_{e\in E(H)} w(e) + w(\hat{M}^*).
\end{align}
From Lemma~\ref{lemma-GreedyMinDist-weight}, $|\hat{M}^*|\geq \lambda|M^*|/(\lambda + 1) \geq 1$.
Let $x = w(\hat{M}^*) / |\hat{M}^*|$. Then, $1\leq \min_{e\in E(H)} w(e) \leq \min_{e\in M^*} w(e) \leq x \leq \max_{e\in E(H)} w(e)$.
Let $\overline{M^*} = M^* \setminus \hat{M}^*$.
From Eq~\eqref{eq-base-weight},
\begingroup
\allowdisplaybreaks
\begin{align} \label{eq-base-ratio}
\frac{w(M)}{w(M^*)} &\leq \frac{|R(M^*)\setminus R(\hat{M})| \max_{e\in E(H)} w(e) + w(\hat{M}^*)}{w(\overline{M}^*) + w(\hat{M}^*)} \nonumber \\
&\leq \frac{(\lambda|M^*| - |\hat{M}^*|)\max_{e\in E(H)} w(e) + |\hat{M}^*|x}{|\overline{M^*}|\min_{e\in M^*} w(e) + |\hat{M}^*|x} \nonumber \\
&\leq \frac{(\lambda|M^*| - |\hat{M}^*|)\max_{e\in E(H)} w(e) + |\hat{M}^*|x}{|\overline{M^*}|\min_{e\in E(H)} w(e) + |\hat{M}^*|x}.
\end{align}
\endgroup
Since $\lambda(|M^*| - |\hat{M}^*|)\max_{e\in E(H)} w(e) \geq |\overline{M^*}|\min_{e\in E(H)} w(e)$ for $\lambda \geq 1$, Eq~\eqref{eq-base-ratio} is non-increasing as $x$ increases.
Taking $x = \min_{e\in E(H)} w(e)$, Eq~\eqref{eq-base-ratio} becomes
\[
\frac{w(M)}{w(M^*)}\leq \frac{\lambda|M^*|\max_{e\in E(H)} w(e)}{|M^*|\min_{e\in E(H)} w(e)} - \frac{|\hat{M}^*|(\max_{e\in E(H)}w(e) - \min_{e\in E(H)} w(e))}{|M^*| \min_{e\in E(H)} w(e)}.
\]
From this and $|\hat{M}^*|(\lambda + 1)/\lambda \geq |M^*|$ (by Lemma~\ref{lemma-GreedyMinDist-weight}),
\begingroup
\allowdisplaybreaks
\begin{align*}
\frac{w(M)}{w(M^*)} &\leq \lambda\mu - \frac{\lambda|\hat{M}^*|(\max_{e\in E(H)}w(e) - \min_{e\in E(H)} w(e))}{|\hat{M}^*|(\lambda + 1)\min_{e\in E(H)} w(e)} \\
&= \lambda\mu - \frac{\lambda(\max_{e\in E(H)}w(e) - \min_{e\in E(H)} w(e))}{(\lambda + 1)\min_{e\in E(H)} w(e)} \\
&= \lambda\mu - \frac{\lambda\cdot \max_{e\in E(H)} w(e)}{(\lambda + 1) \min_{e\in E(H)} w(e)} + \frac{\lambda\cdot \min_{e\in E(H)} w(e)}{(\lambda + 1) min_{e\in E(H)} w(e)} \\
&= \frac{\lambda^2\mu + \lambda\mu - \lambda\mu + \lambda}{\lambda + 1} = \frac{\lambda^2\mu + \lambda}{\lambda + 1}.
\end{align*}
\endgroup
If $M = M_2$, $\frac{w(M_2)}{w(M^*)} \leq \frac{\lambda^2\mu + \lambda}{\lambda + 1}$ since $w(M_2) \leq w(M_1)$.
Therefore, the theorem holds.
\end{proof}

\subsection{Algorithm GreedyMinNum}
Let $H(U\cup V,E,w)$ be the hypergraph constructed for the an instance of the MTRS-minNum problem.
%Algorithm GreedyMinNum selects an edge containing the most number of uncovered riders in each iteration.
\begin{itemize}
\item Initially, $M=\emptyset$. In each iteration, select $e$ in $E(H)$ with maximum $|R(e)|$, $M=M\cup \{e\}$, and remove all vertices of $e$ and all edges of $\cup_{u\in e}E(u)$. Continue to the next iteration until $R(M) = R$.
\end{itemize}
Algorithm GreedyMinNum has at most $|R|$ iterations to compute $M$.
By a similar analysis of Lemma~\ref{lemma-Greedy-feasible}, Algorithm GreedyMinNum computes a feasible solution $M$ to the MTRS-minNum problem. We show that GreedyMinNum is $\frac{\lambda + 2}{2}$-approximate if Assumption~\ref{assumption-designated-drivers-strong} is satisfied.
Let $M = \{e_1,\ldots, e_{|M|}\}$ be the solution found by GreedyMinNum and $M^*= \{e^*_1,\ldots,e^*_z\}$ be an optimal solution to the MTRS-minNum problem.
For any $1\leq i \leq |M|$, let $M_i = \bigcup_{1 \leq a \leq i} e_a$ with $M_0 = \emptyset$ and $D(M_0) = R(M_0) = \emptyset$.
For any $1\leq i \leq z$, let $DM^*_i = \{e^* \in M^*_i \mid D(e^*) \in D(M_i)\}$ with $DM_0 = \emptyset$.

%Let $M = \{e_1, e_2,\ldots, e_{|M|}\}$ be the solution found by GreedyMinNum, where $e_i$ is the $i^{th}$ edge selected and added to $M$, $1 \leq i \leq |M|$ (at the $i^{th}$ iteration).
%Note that the number of covered drivers is $|M|$, that is, $|D(M)| = |M|$.

%We now show that GreedyMinNum is $\frac{\lambda + 2}{2}$-approximate if Assumption~\ref{assumption-designated-drivers-strong} is satisfied.
%Let $M^*$ be an optimal solution to an instance of the MTRS-minNum problem, and we show that $\frac{|M|}{|M^*|} \leq \frac{\lambda + 2}{2}$.
%For analysis purpose, we label the edges of $M^*$ as $M^* = \{e^*_1,e^*_2,\ldots,e^*_z\}$, where $z = |M^*|$.
%For any $1\leq i \leq z$, let $DM^*_i = \{e^* \in M^*_i \mid D(e^*) \in D(M_i)\}$ with $DM_0 = \emptyset$.
%Note that $z \leq |M|$.

\begin{lemma} \label{minNum-Greedy-property}
For each edge $e_i$ in $M_z$, $|R(e_i)| \geq |R(e^*) \setminus R(M_{i-1})|$ for every edge $e^* \in M^* \setminus DM^*_{i-1}$.
\end{lemma}

\begin{proof}
At the beginning of each iteration $i$ of the algorithm, $1 \leq i \leq z$, $M^* \setminus DM^*_i \neq \emptyset$.
In the $i^{th}$ iteration, the algorithm selects an edge $e_i$ with the maximum $|R(e_i) \setminus R(M_{i-1})|$.
Notice that if $|R(e^*)\setminus R(M_{i-1})| = 0$ for every edge $e^* \in M^* \setminus DM^*_i$, $|R(e_i)| > |R(e^*)\setminus R(M_{i-1})|$, implying the lemma.
Assume there is a set of edges $E^* \subseteq M^* \setminus DM^*_i$ such that for every edge $e^*\in E^*$, $R(e^*)\setminus R(M_{i-1}) \neq \emptyset$.
For every edge $e^* \in E^*$, there is an edge $e \in E(H)$ such that $D(e) = D(e^*)$ and $R(e) = R(e^*)\setminus R(M_{i-1})$ by Observation~\ref{obs-1}.
Since $e$ is not incident to any edge of $M_{i-1}$, $e$ is a candidate for the algorithm to select.
Since the algorithm selects $e_i$, $|R(e_i)| \geq |R(e)| = |R(e^*_j)\setminus R(M_{i-1})|$.
\end{proof}

\begin{theorem} \label{minNum-Greedy-theorem-app}
Let $M$ be a solution found by Algorithm GreedyMinNum and $M^*$ be an optimal solution to the MTRS-minNum problem. Then $\frac{|M|}{|M^*|} \leq \frac{\lambda + 2}{2}$ for $\lambda \geq 2$, under Assumption~\ref{assumption-designated-drivers-strong}.
\end{theorem}

\begin{proof}
By Lemma~\ref{minNum-Greedy-property}, in each iteration $i$ of the first $z = |M^*|$ iterations, the algorithm selects an edge $e_i$ with $|R(e_i)| \geq |R(e^*_j) \setminus R(M_{i-1})|$, for every $e^* \in M^* \setminus DM^*_{i-1}$.
At most one edge $e^* \in M^* \setminus DM^*_{i-1}$ is included in $DM^*_i$ by definition, and if one such edge $e^*$ is included in $DM^*_i$, then $e^*$ will not be considered in the $j^{th}$ iteration of the algorithm to select $e^*_j$, for $i < j \leq z$ (as $D(e^*) \in M_i$).
Hence, at most $|R(e^*) \setminus R(M_{i-1})| \leq |R(e_i)|$ riders would not be covered by $M_i$.
From this, the total number $x$ of riders in $R(M^*)$ that are not covered by $M_z$ is at most $|R(M_z)|$, implying $x \leq |R(M^*)|/2$ in the worst case.
Since $|R(e^*)| \leq \lambda$, $|R(M^*)| \leq \lambda|M^*|$.
To cover $|R(M^*)|/2$ riders by edges of $M \setminus M_z$ requires at most $\lambda|M^*|/2$ edges in $M \setminus M_z$.
From this and $|M_z| = |M^*|$, we obtain
\begin{align*}
|M| \leq |M^*| + \frac{\lambda|M^*|}{2} = \frac{2|M^*| + \lambda|M^*|}{2} = \frac{(\lambda + 2)|M^*|}{2}.
\end{align*}
Hence, $\frac{|M|}{|M^*|} \leq \frac{\lambda + 2}{2}$.
\end{proof}

\subsection{Algorithm LS}
Based on Assumption~\ref{assumption-designated-drivers-strong}, which is stronger than Assumption~\ref{assumption-drivers}, GreedyMinDist and GreedyMinNum achieve better approximation ratios than the trivial upper bounds, namely $\lambda\mu$ for MTRS-minDist and $\lambda$ for MTRS-minNum with Assumption~\ref{assumption-drivers}.
We give a local search algorithm (Algorithm LS) for MTRS-minNum with Assumption~\ref{assumption-drivers} only.

Let $H(U \cup V,E,w)$ be the hypergraph constructed for an instance of the MTRS-minNum problem.
Recall that $U(H) \subseteq D(H) = \Delta(H)$ for MTRS-minNum.
Algorithm LS has two steps. In Step (1), it finds a feasible solution $M' \subseteq E(H)$, and in Step (2), it improves $M'$ to obtain a final solution $M \subseteq E(H)$.
Below is a more detailed description:
\begin{itemize}
\item[(1)] Algorithm LS starts Step (1). Let $H'$ be the subgraph of $H$ with all hyperedges removed. Find a maximum cardinality matching $M'$ in $H'$.
\item[(2)] Algorithm LS starts Step (2). Let $M = M'$. In each iteration, the algorithm tries to find an improvement $\delta_e$ of an edge $e \in M$.
A set $\delta_e$ of pairwise disjoint edges in $N_H(e)$ is an \emph{improvement} if $M = (M \setminus N_H(\delta_e)) \cup \delta_e$ results in a new solution $M$ such that $R(M) = R$, $|D(M)|$ is decreased by at least one, and all edges of $M$ are pairwise vertex-disjoint. In other words, $M = (M \setminus N_H(\delta_e)) \cup \delta_e$ is a new solution by adding $\delta_e$ to $M$ and removing all edges of $M$ that are incident to $\delta_e$. %An improvement $\delta_e$ is \emph{maximal} if $|D(M)|$ decreased the most among all improvements of $e$.
	\begin{itemize}
	\item For any edge $e \in M \cap M'$ (equivalently, $e\in M(1) = \{e \in M \mid |R(e)| = 1\}$), find an improvement $\delta_e$. Perform the augmentation $M = (M \cup \delta_e) \setminus N_H(\delta_e)$, and repeat this for another edge in $M \cap M'$ until no improvement can be found or $M \cap M' = \emptyset$ (equivalently, $M(1) = \emptyset$).
	\end{itemize}
\end{itemize}

\begin{figure}[!ht]
\centering
\includegraphics[width=.9\linewidth]{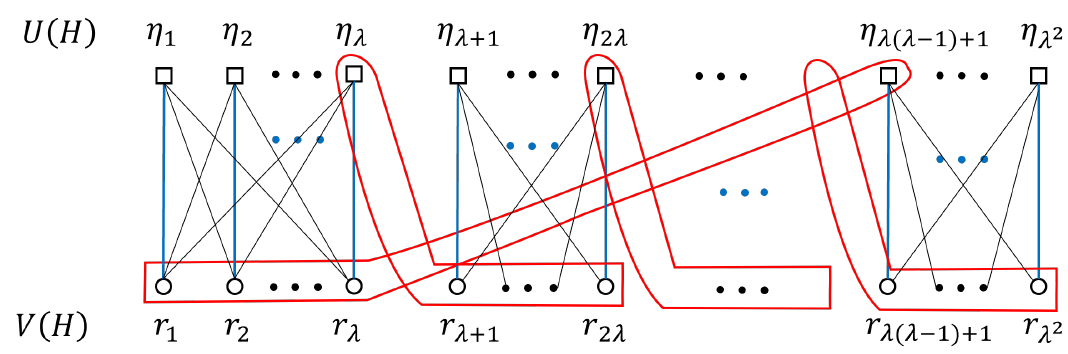}
\caption{The hypergraph $H(U\cup V,E,w)$ with $k = \lambda$. The blue edges belong the solution found by Algorithm LS, the red edges belong to an optimal solution, and black edges are the rest of edges in $H$.
Since there are $\lambda^2$ blue edges and $\lambda$ red edges, the approximation ratio is $\lambda$.}
\label{fig-LS-Example}
\end{figure}

In the worst case, Algorithm LS has an approximation ratio of $\lambda$ as shown in Figure~\ref{fig-LS-Example}, and we called Algorithm LS a heuristic.

\section{Clustering} \label{sec-clustering}
Computing all feasible matches and the hypergraph $H(U\cup V,E,w)$ can be time-consuming when the problem instance size is large. Furthermore, a designated driver may need to travel a long distance to serve a rider in order to satisfy Assumption~\ref{assumption-designated-drivers-strong}.
To address these issues, we propose a clustering algorithm (called \textbf{CL}) to partition the drivers and riders into groups, where each group contains a limited number of drivers and riders.
Given the set $\Gamma$ of personal drivers and the set $R$ of riders in an instance of MTRS, $\Gamma \cup R$ is partitioned into disjoint groups $\mathcal{C} = \{C_1,C_2,\ldots,C_l\}$ such that 
$C_i \cap C_j = \emptyset$ for all $i \neq j$ and $\cup_{1\leq i \leq l} C_i = \Gamma \cup R$. Each $C_i$ is 
called a \emph{cluster}, and denoted by $\Gamma(C_i)$ and $R(C_i)$ are the sets of drivers and riders contained in $C_i$, respectively. We then allocate designated drivers to clusters. For each cluster, the feasible matches and hypergraph are computed independently.

Next, we introduce some definitions and notation that are used in Algorithm CL.
Recall that the travel parameters of each driver and rider are summarized in Table~\ref{table-problem-notation}, so they are not re-introduced here.
In Algorithm CL, each driver $\eta_u \in \Gamma$ is represented by a \textit{driver object} $\eta_u = \{o_u, d_u, [a_u, b_u]\}$ of three features/attributes, where $o_u$ is the origin and $d_u$ is the destination  of driver $\eta_u$, and $[a_u, b_u]$ is the driver \textit{interval window}, where both are input parameters to the clustering algorithm.
Similarly, each rider $r_v \in R$ is represented by a \textit{rider object} $r_v = \{o_v, d_v, [a_v, b_v]\}$ of three attributes, where $[a_v, b_v]$ is the rider interval window.
The interval window is a sub-interval between the driver's (rider's) earliest departure time and latest arrival time.
When two intervals of two objects overlap, it suggests that their time window can be compatible to each other, so interval windows are used in determining whether two objects are connected.
In our algorithm, for any FM-driver $\eta_u$, $a_u = \alpha_u$ and $b_u = \beta_u - z_u$, and for any LM-driver $\eta_u$, $a_u = \alpha_u + 0.2 \cdot z_u$ and $b_u = \beta_u - 0.3 \cdot z_u$, where $z_u$ is the maximum detour time for $\eta_u$.
For any FM-rider $r_v$, $a_v = \alpha_v$ and $b_v = \beta_v - \theta_v\cdot \hat{t}(o_v, d_v)$, and for any LM-rider $r_v$, $a_v = \alpha_v + 0.25 \cdot \theta_v \cdot \hat{t}(o_v, d_v)$ and $b_u = \beta_v - 0.35 \cdot \theta_v \cdot \hat{t}(o_v, d_v)$, where $\hat{t}(o_v, d_v)$ is the travel duration of the fastest public transit from $o_v$ to $d_v$.
For a set $\Gamma' \subseteq \Gamma$, $\bar{a}(\Gamma') = \frac{\sum_{\eta_u \in \Gamma'} (a_u)}{|\Gamma'|}$ and $\bar{b}(\Gamma') = \frac{\sum_{\eta_u \in \Gamma'} (b_u)}{|\Gamma'|}$ are the average driver start and end intervals of $\Gamma'$, respectively.
Similarly, for a set $R' \subseteq R$, $\bar{a}(R')$ and $\bar{b}(R')$ are the average driver start and end intervals of $R'$, respectively.
Let $\Gamma(FM)$ and $\Gamma(LM)$ be the sets of personal FM-drivers and LM-drivers, resp.
Let $R(FM)$ and $R(LM)$ be the sets of FM-riders and LM-riders, resp.
Algorithm CL has two phases, described in the following.

\subsection{Construction phase (I)} \label{subsec-construct-clusters}
Clusters are created for FM match type first, and then for LM match type, which are independent of each other.
There are three major steps: (A) divide the road-map into smaller cells, (B) distribute the drivers of $\Gamma(FM)$ and riders of $R(FM)$ to cells according to their origins and create clusters for these drivers and rides in each cell, and (C) repeat step (B) for $\Gamma(FM)$ and $R(FM)$.

\begin{figure}[!ht]
	\centering
	\includegraphics[width=.55\linewidth]{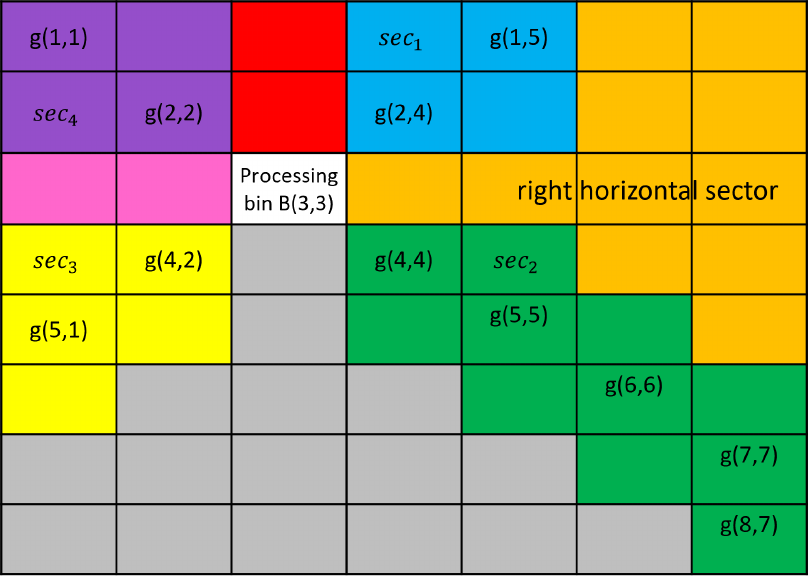}
	\caption{The road-map is viewed as 8 sectors of cells, centered at $g(3,3)$.
	The blue area is the diagonal sector $sec_1$, the green area is the diagonal sector $sec_2$, and the orange area is the right horizontal sector, and so on.}
	\label{fig-FM-cluster-sector}
	\end{figure}

(I.A) Typically, the spatial area (physical road-map) can be viewed as a large rectangle.
We partition the large rectangle into $m_1\times m_2$ smaller rectangles called \emph{cells}, where $1\leq m_1$ and $1\leq m_2$ are input parameters (integers) to Algorithm CL. Figure~\ref{fig-FM-cluster-sector} shows a rectangle is divided into $8\times 8$ cells.
Each cell is denoted by $g(x,y)$ for $1\leq x \leq m_1$ and $1 \leq y \leq m_2$, where $g(1,1)$ is the top left corner. Two cells $g(x,y)$ and $g(x',y')$ are \emph{adjacent} if $|x - x'| + |y - y'| \leq 2$ (at most two cells away). For each cell $g(x,y)$, we create a bin $B(x,y)$.

(I.B1) Distribute riders of $R(FM)$ and drivers of $\Gamma(FM)$ to each bin $B(x,y)$ if their origins are in cell $g(x,y)$. Let $R(x,y)$ and $\Gamma(x,y)$ be the sets of riders and drivers in bin $B(x,y)$, respectively.
For each cell $g(x,y)$, the $m_1\times m_2$ cells are divided into 8 non-overlapping sectors/areas, 4 diagonal sectors (labeled as $sec_1$, $sec_2$, $sec_3$ and $sec_4$), 2 horizontal and 2 vertical sectors (labeled as $sec_5$, $sec_6$, $sec_7$ and $sec_8$), all centered at $g(x,y)$.
	An example is shown in Figure~\ref{fig-FM-cluster-sector}.

(I.B2) Riders and drivers with similar travel direction (determined by their bins and sectors) and overlapping time windows (indicating compatible travel time) are put into the same cluster. Let $R(sec_j) \subseteq R(x,y)$ be the set of riders and $\Gamma(sec_j) \subseteq \Gamma(x,y)$ be the set of drivers with destinations in sector $sec_j$, $1\leq j \leq 8$, respectively. For each $1 \leq j \leq 8$, find a subset $R'_j$ of $R(sec_j)$ such that for every $r_v \in R(sec_j) \setminus R_j'$, $[a_v, b_v]$ overlaps with $[a_u, b_u]$ for some rider $r_u \in R(sec_j)$.
This can be done by using a heuristic to find a dominating set $R'_j$ in the interval graph formed from the interval windows of riders in $R(sec_j)$. 
Let $C_o$ be the origin cell and $C_d$ be the destination sector for cluster $C$. Then, clusters are created as follows.
\begin{itemize}
	\item For each rider $r_u \in R'_j$, create a FM-cluster $C(r_u)$ with $C_o = g(x,y)$ and $C_{d} = sec_j$, and add $r_u$ to $C(r_u)$. To add riders and drivers to $C(r_u)$, we first construct an interval graph $I_j(U\cup V,E)$ as follows. $U(I_j) = R'_j$ and $V(I_j) = (R(sec_j) \setminus R_j') \cup \Gamma(sec_j)$. For any $u \in U(I_j)$ and $v \cup V(I_j)$, add an edge $\{u,v\}$ to $E(I_j)$ if $[a_u, b_u]$ and $[a_v, b_v]$ overlap. For every node $u$ in $I_j$, let $\deg_{I_j}(u)$ be the degree of $u$.
	\item For every rider node and driver node $v \in V(I_j)$ with $\deg(v) = 1$, add the rider/driver corresponding to $v$ to the corresponding cluster $C(r_u)$, where $\{u,v\} \in N(u)$, and remove $v$ from $I_j$. Then in each iteration, select the cluster $C(r_u)$ with the smallest $|R(C(r_u))|$ ($\deg_{I_j}(u)$ as a tiebreaker). Pick the node $v$ adjacent to $u$ with smallest $\deg_{I_j}(v)$, add $v$ to $C(r_u)$, and remove $v$ from $I_j$. Repeat this until all riders and drivers are clustered. For every isolated node $v$ of $V(I_j)$, create a cluster $C(\eta_v)$, which will be used in the refinement phase.
\end{itemize}

(I.C) Repeat Step (I.B) symmetrically for $R(LM)$ and $D(LM)$ (i.e., treat destinations as origins and vice versa). Distribute riders of $R(LM)$ and drivers of $\Gamma(LM)$ to each bin $B(x,y)$ if their destinations are in cell $g(x,y)$. Then process each bin $B(x,y)$ independent of other bins.
Construct the sectors as described in Step (I.B1).
Define $R(sec_j) \subseteq R(x,y)$ and $\Gamma(sec_j) \subseteq \Gamma(x,y)$ to be the sets of riders and drivers with origins in sector $sec_j$, respectively.
Then repeat the rest of Step (I.B2) to create LM-clusters for $R(sec_j)$ and $\Gamma(sec_j)$ for each $1 \leq j \leq 8$.

\subsection{Refinement phase (II)} \label{subsec-adjust-clusters}
The input to the second phase of Algorithm CL is the set $\mathcal{C}$ of clusters constructed in phase (I).
For brevity, let $|C| = |R(C) \cup \Gamma(C)|$ be the \emph{size} of $C$ for any cluster $C$ of $\mathcal{C}$.
Large clusters are divided into multiple smaller clusters, while small clusters are merged with nearby clusters.
Further, we try to keep $\frac{|R(C)|}{|\Gamma(C)|}$ to be within some reasonable range for each cluster $C \in \mathcal{C}$.

(II.A) First, small clusters are merged with their adjacent clusters (small adjacent clusters have a higher priority).
Let $s_{min}$ be an input parameter to Algorithm CL (denoting the threshold of small cluster size) and $C_{min} = \{C \in \mathcal{C} \mid |C| < s_{min}\}$.
Construct a graph $G_C(U \cup V,E)$ for clusters of $\mathcal{C}$, where $U(G_C) = C_{min}$ and $V(G_C) = \mathcal{C} \setminus C_{min}$.
Two clusters $C$ and $C'$ are \emph{adjacent} if $C_{d} = C'_{d}$, $C_o$ and $C'_o$ are adjacent, and $[\bar{a}(\Gamma(C)), \bar{b}(\Gamma(C))]$ overlaps with $[\bar{a}(R(C')), \bar{b}(R(C'))]$ or $[\bar{a}(R(C)), \bar{b}(R(C))]$ overlaps with $[\bar{a}(\Gamma(C')), \bar{b}(\Gamma(C'))]$.
For every pair $C \in U(G_C)$ and $C' \in V(G_C)$, add an edge $\{C, C'\}$ to $E(G_C)$ if $C$ and $C'$ are adjacent.
For every pair $C, C' \in U(G_C)$ with $C \neq C'$, add an edge $\{C, C'\}$ to $E(G_C)$ if $C$ and $C'$ are adjacent.

\begin{itemize}
\item Process each cluster $C$ of $C_{min}$, starting with the one that has the smallest size.
Select a cluster $C'$ in $G_C$ with the smallest size that is adjacent to $C$ (if there is none, remove $C$ from $C_{min}$).
\textit{Merge} clusters $C$ with $C'$ by moving all riders and all drivers of $C$ to $C'$, and remove $C$ from $C_{min}$.
If $C'$ was in $C_{min}$ before the merge and $|C'|$ becomes at least $s_{min}$ after the merge, remove $C'$ from $C_{min}$.
Contract the two vertices corresponding to $C$ and $C'$ in $G_C$.
Repeat until $C_{min}$ is empty, which is the end of this step.
\end{itemize}

(II.B) Given two input parameters $\tau_1$ and $\tau_2$ to Algorithm CL, where $0 < \tau_1 \leq |R|/|\Gamma|$ and $\tau_2 \geq 1$, a cluster $C \in \mathcal{C}$ is \emph{balanced} if $\tau_1 \leq \frac{|R(C)|}{|\Gamma(C)|} \leq \frac{\tau_2|R|}{|\Gamma|}$; otherwise, $C$ is \emph{imbalanced}.
Let $C_{imb} = C_l \cup C_h$, where $C_{l} = \{C \in \mathcal{C} \mid \frac{|R(C)|}{|\Gamma(C)|} < \tau_1\}$ and $C_{h} = \{\frac{|R(C)|}{|\Gamma(C)|} > \frac{\tau_2|R|}{|\Gamma|}\}$ for the resulting set $\mathcal{C}$ of clusters after Step (II.A).
For each cluster $C \in C_{imb}$, define an \emph{imbalanced value} $f_C$ to indicate how imbalanced $C$ is:
\begin{equation*}
f_C = 
	\begin{cases}
	\frac{|\Gamma(C)|}{|R(C)|}, & \text{if } C \in C_l, \\
	\frac{|R(C)|}{|\Gamma(C)|}, & \text{if } C \in C_h.
	\end{cases}
\end{equation*}
%Sort $C_{imb}$ in decreasing order of the clusters' imbalanced values.
Construct a graph $G_C(U \cup V,E)$ for clusters of $\mathcal{C}$, where $U(G_C) = C_{imb}$ and $V(G_C) = \mathcal{C} \setminus C_{imb}$.
Create edges for clusters between $U(G_C)$ and $C' \in V(G_C)$ and edges for clusters between $U(G_C)$ and $U(G_C)$ as described in Step (II.A).
\begin{itemize}
\item Process each cluster $C$ of $C_{imb}$, starting with the one that is most imbalanced.
Select a cluster $C'$ in $G_C$ that is adjacent to $C$ (if there is none, remove $C$ from $C_{imb}$) such that merging $C$ with $C'$ reduces $f_C$ the most (e.g., if $C \in C_l$, selects the cluster $C'$ from $C_h$ with the highest $f_{C'}$).
Merge cluster $C$ with $C'$, and remove $C$ from $C_{imb}$.
Contract the two vertices corresponding to $C$ and $C'$ in $G_C$.
Repeat until $C_{imb}$ is empty, which is the end of this step.
\end{itemize}

(II.C) Finally, large clusters are divided.
Let $s_{max}$ be an input parameter to Algorithm CL (denoting the threshold of large cluster size), $\mathcal{C}$ be the set of clusters obtained after Step (II.B), and $C_{max} = \{C \in \mathcal{C} \mid |C| > s_{max}\}$.
The algorithm processes each $C \in C_{max}$ one by one.
$C$ is heuristically partitioned into $C_{1}, C_{2},\ldots, C_{z}$, where $z = \floor{|C_i|/s_{max}}$, such that $\sum_{1\leq i \leq z} |R(C_i)| = |R(C)|$, $\sum_{1\leq i \leq z} |\Gamma(C_i)| = |\Gamma(C)|$, $|C_i| \approx |C|/z$ and $|R(C_i)|/|\Gamma(C_i)| \approx |R(C)|/|\Gamma(C)|$ for every $1 \leq j \leq z$.
After the partitioning, $C_{1},\ldots, C_{z}$ are included into $\mathcal{C}$, and they do not need to be checked again.

\paragraph{Clustering the designated drivers.}
If Assumption~\ref{assumption-drivers} is known before the clustering process, we can simply assign each distinct designated driver $\eta_u \in \Delta$ into the cluster containing the rider $r_v$ that can be served by $\eta_u$ after Step (II.C).
In the case where we only know $|\Delta| = |R|$, after Step (II.C), we can heuristically assign each rider $r_v$ to a designated driver $\eta_u$ based on the distance between $o_u$ and $o_v$ for FM match type or between $d_u$ and $d_v$ for LM match type.
For example, randomly select an unmarked driver $\eta_u \in \Delta$ and pair it with $r_v$ where $\dist(o_u, o_v)$ is the shortest among all unmarked riders of $R$. Then, mark $\eta_u$ and $r_v$, and select the next unmarked driver until all are marked.
A more accurate approach is to compute a complete bipartite graph between $\Delta$ and $R$, where the weight of each edge $\{\eta_u, r_v\}$ is $\dist(o_u, o_v)$ for FM or $\dist(d_u, d_v)$ for LM.
Then, find a minimum weight perfect matching $M$ on this complete bipartite graph.
Each rider is associated with a distinct driver w.r.t. $M$, so designated drivers can be partitioned based on $M$.

\section{Experiment} \label{sec-experiment}
We conduct an empirical study to evaluate our model and algorithms for the centralized transit system that integrates public transit and ridesharing, for FM/LM-transit specifically.
Based on a number of real-life datasets and surveys related to transit and travel patterns in Chicago City, we create a simulation environment where rider requests and driver offers are generated and continuously processed by the system.

\subsection{Simulation setup and datasets overview}
The road-map data of Chicago City is obtained from OpenStreetMap (BBBike.org)\footnote{Last access, May 2024. Planet OSM. BBBike. \url{https://download.bbbike.org/osm}}.
The bounding box of the road-map is: bottom-left (41.6550669, -87.9379308) and top-right (42.0273554, -87.5453074).
The logical graph data structure of the road-map is constructed using GraphHopper\footnote{GraphHopper 9.0. \url{https://www.graphhopper.com}}, which contains 756,661 edges and 507,199 vertices.
The schedule-based transit network is also constructed using GraphHopper with the GTFS data obtained from Chicago Transit Authority (CTA)\footnote{CTA. \url{https://www.transitchicago.com}. CDP. \url{https://data.cityofchicago.org}\label{fn1}}.

The Chicago City is divided into 77 official community areas (\textbf{area} for brevity and labeled as
A1 to A77).
We examined a transit ridership dataset/report (labeled as \textit{PTD}) maintained by CTA, travel survey 2019 results conducted by CMAP\footnote{Chicago Metropolitan Agency for Planning(CMAP). \url{https://www.cmap.illinois.gov}}, a transit-trend report by CMAP, and a rail study~\cite{CTA19} in Chicago community areas to reveal some rider travel patterns. We specifically examined pre-pandemic data if available to observe general trends.
The PTD dataset contains bus and rail ridership (which shows the monthly averages and monthly totals of August 2019 for all CTA bus routes and Chicago `L' rail station entries), and it is obtained from Chicago Data Portal (CDP) and CTA~\footref{fn1}.
The travel survey 2019, known as \textit{My Daily Travel 2019}, was conducted from late 2017 to mid 2019 for northeastern Illinois, involving more than 12000 households~\cite{CMAP-MDT19}. Within the survey results, we are particularly interested in the transit dataset (labeled as \textit{CMAP-PT}).
In the CMAP-PT dataset, a \emph{public transit trip} consists of an origin and a destination (given in latitude and longitude), departure and arrival times, and a sequence of transfers.
We include all public transit trips that use CTA transit where origins or destinations are within Chicago, which results in 7900 \textit{CTA-trips}.
Due to some of these trips have one or more transfers, the total transit ridership is higher than 7900.
We generate riders based on CMAP-PT.

All the busiest bus routes in Chicago City visit some Chicago `L' stations; and in fact, many of the bus routes visit multiple stations.
Using all CTA-trips where the origin and destination of each trip is within Chicago, we calculated the total numbers of areas that contain the origins (denoted as \textit{departure areas}) and the destinations (denoted as \textit{arrival areas}), respectively.
Riders generated will have origins and destinations from the most common visited areas.
According to CMAP, more than 68\% of all public transit trips in the Chicago region are work commutes.
According to Moovit\footnote{\url{https://moovitapp.com/insights/en/Moovit_Insights_Public_Transit_Index-countries.}}, about 1/3 is short commute (up to 30 minutes), 1/3 is medium commute (30 - 60 minutes) and 1/3 is long commute (60+ minutes).

From American Community Survey 2019 for Chicago City, 55.4\% of respondents drive to work, with 48.2\% drive alone and 7.1\% carpooled (5.2\% in 2-person carpool); and about 28.4\% of people use public transportation for work commute.
\begin{table}[!ht]
\parbox{.53\linewidth}{
\centering
	\begin{tabular}{c | l}
   	Time of departure to work & Percent of people \\ \hline
    6:00 am to 6:59 am & 15.1\% \\
	7:00 am to 7:29 am 	& 14.6\% \\
    7:30 am to 7:59 am	& 10.1\% \\
	8:00 am to 8:29 am	& 14.8\% \\
    8:30 am to 8:59 am	& 7.8 \% \\
    9:00 am to 11:59 pm &25.1\% \\
   	\hline
   \end{tabular}
}
%\hfill
\parbox{.46\linewidth}{
\centering
	\begin{tabular}{c | l}
   	Travel time to work  & Percent of people \\ \hline
    < 15 minutes &	10.4\% \\
    15 to 24 minutes	  & 20.7\% \\
    25 to 34 minutes	  & 25.4\% \\
    35 to 44 minutes	  & 11.3\% \\
    45 to 59 minutes	  & 16.2\% \\
    60 or more minutes	  & 16.0\% \\
   	\hline
   \end{tabular}
}
\label{table-ACS}
\caption{Basic stats of drive commuters in the American Community Survey 2019}
\end{table}
This indicates there are many personal drivers and at least a portion of them are willing to provide ridesharing service.
The morning-drive peak hours (shown in Table~\ref{table-ACS}, along with drive duration) mostly align with the findings in My Daily Travel 2019, similarly for the drive and transit peak hours.

We also examined ridesharing datasets in 2018/2019, reported by Transportation Network Providers (which are rideshare companies), and analyzed their results. The whole ridesharing dataset is publicly available at Chicago Data Portal
The number of people that use ride-hailing/ridesharing in Chicago has been increasing, with about 30\% reporting they use these services for commuting to work~\cite{CMAP-TT} (the 30\% is for the CMAP region, so it is likely higher for just Chicago City).
The average number of ridesharing trips per hour is more than 13000 during morning peak hours and 15000-25000 during afternoon peak hours~\cite{CMAP-TNP}.
This is similar to the finding in~\cite{Gu-COR24} for longer duration/distance ridesharing trips (beside the airport areas A77 and A64/A56).
Based on the studies/reports above, we generate trips in the morning and afternoon peak hours as test instances for the experiments.

\subsubsection{Instance generation}
Since the majority of frequent transit routes pass through multiple metro stops/stations, the set $TS$ of fixed transit stations in any instance of MTRS is the set of all metro (Chicago `L') stations within Chicago.
We select two \textit{time periods} as a representation of the major traffic patterns that are the focus of this paper: \textbf{morning peak} and \textbf{afternoon peak} periods.
Each time period consists of consecutive fixed time intervals (each interval is 30 minutes).
The morning peak period consists of 3 time intervals from 6:00 am to 7:30 am and afternoon peak period consists of 5 time intervals from 3:00 pm to 5:30 pm.
At the start of each time interval, ridesharing requests and offers are accumulated during the interval, and then all requests and offers are processed together at the end of that interval - forming the MTRS instance for that interval.
Around 800-1000 and 900-1200 riders $R$ are generated for each time interval in the morning peak period and afternoon peak period, respectively.
The number of designated drivers generated equals the number of the riders generated, and the number of personal drivers generated is about 1/3 of the riders generated. 
Detailed description of riders' and drivers' parameters is given below (a more detailed description can be found in the appendix).

\subsubsection{Generation of riders} Let $[t_a, t_b)$ be a time interval (e.g., 6:00 am to 6:29:59 am).
For each rider $r_j \in R$ generated during $[t_a, t_b]$, its earliest departure time $\alpha_j$ is selected uniformly at random from $[t_a + 45 \text{ minutes}, t_a + 75 \text{ minutes}]$.
The latest arrival time $\beta_j$ of a rider $r_j$ is selected uniformly at random from $[\alpha_j + 45 \text{ minutes}, \alpha_j + 90 \text{ minutes}]$.
Every rider has the same acceptance threshold $\theta_j = 0.3$ (save at least 30\% of transit time).
To assign an origin $o_j$ to rider $r_j$, we use only the most common departure areas based on the CMAP-PT dataset. 
For each time interval $[t_a, t_b)$, let $DA(t_a,t_b)$ and $AA(t_a,t_b)$ be the sets of the most common departure areas and arrival areas of the transit trips in CMAP-PT with departure time during interval $[t_a, t_b)$, respectively.
Suppose $\alpha_j$ (decided above) falls in the interval $DA(t_{a'},t_{b'})$.
First, a departure area $DA_i$, $1\leq i \leq 15$, of $DA(t_{a'},t_{b'})$ is selected with probability $n(DA_i) / n(DA(t_{a'},t_{b'}))$, where $n(DA_i)$ is the number of departures in area $DA_i$ and $n(DA(t_{a'},t_{b'}))$ is the total number of departures in $DA(t_a,t_b)$.
Then, a point (latitude, longitude) in area $DA_i$ is selected uniformly at random to be assigned as the origin $o_j$ for rider $r_j$.
The destination $d_j$ for rider $r_j$ is decided exactly the same way symmetrically.
When generating a rider $r_j$, we use GraphHopper to compute the fastest public transit route $\pi(r_j)$ from $o_j$ to $d_j$ with departure time $\alpha_j$ (decided as above).
If $r_j$ cannot reach $d_j$ before $\beta_j$ using $\pi(r_j)$, travel duration of $\pi(r_j)$ is less than 30 minutes, or the total walking distance of $\pi(r_j)$ exceeds 2.5 miles $\approx$ 4 km, then $\pi(r_j)$ is considered \textit{invalid}, $r_j$ is discarded, and a new $r_j$ is generated randomly until a valid $\pi(r_j)$ is obtained. 

\subsubsection{Generation of drivers} 
After the set $R$ of riders has been generated for interval $[t_a, t_b]$, we generate the set $\Gamma$ of personal drivers.
For a departure area $A_i \in DA(t_a,t_b)$, let $R_{FM}(A_i)$ be the sets of generated FM-riders with origins in $A_i$.
$|R_{FM}(A_i)|/3$ personal FM-drivers are generated with random origins in $A_i$, denoted as $D_{FM}(A_i)$.
For each driver $\eta_i \in D_{FM}(A_i)$, the destination of $\eta_i$ is decided in the same way as how the destination of a rider is selected.
Similarly, $D_{LM}(A_i)$ is the set of personal LM-drivers generated with random destinations in $A_i$, based on $R_{LM}(A_i)$.
Then, $\Gamma = \cup_{\forall A_i} D_{FM}(A_i) \cup D_{LM}(A_i)$ becomes the set of personal drivers for interval $[t_a, t_b]$.
The maximum detour $z_i$ of $\eta_i \in \Gamma$ as at least 20 minutes to at most $t(FP(o_i, d_i))$, where $t(FP(o_i, d_i))$ is the time needed to traverse the shortest path from $o_i$ to $d_i$.
The vehicle capacity $\lambda_i$ is selected uniformly at random from [2, 3].
The earliest departure time $\alpha_i$ is between $[t_a + 30 \text{ minutes}, t_a + 70 \text{ minutes}]$ for both personal and designated drivers, and latest arrival time $\beta_i$ is between $[\alpha_i + 45 \text{ minutes}, \alpha_i + 70 \text{ minutes}]$ for any generated personal driver $\eta_i$.
After each rider $r_j$ is generated during interval $[t_a, t_b]$, a designated driver $\eta_i \in \Delta$ will be generated. For a FM-rider $r_j$, $\eta_i$ is a FM-driver with origin $o_i$ randomly generated in the same area of the origin of $r_j$; otherwise, $\eta_i$ is a LM-driver with origin $o_i$ randomly generated in the same area of the destination of $r_j$.
In our experiment, destination $d_i = o_i$ for every $\eta_i$, and detour limit is ignored for designated drivers.
The vehicle capacity $\lambda_i$ is 3 for all designated drivers $\eta_i$.

\subsubsection*{Computing feasible matches}
The simulation and all algorithms were implemented in Java, and the experiment were conducted on an AMD Ryzen 5 7600 processor with 20 GB RAM allocated to JVM, using up to 8 threads.
All ILP formulations in our algorithms were solved using CPLEX v12.10.
Only the road-map data structure and the transit network were pre-computed in the beginning.
Any other computation, including shortest paths and fastest transit routes computation, was done in real-time for each time interval.
After riders and drivers are generated, we compute the feasible matches and construct the hypergraph $H$, as introduced in Section~\ref{subsec-exact}.
If clustering is not used, the feasible matches are computed between every driver in $D$ and every rider in $R$; otherwise, between every rider and every rider within the same cluster.
%It is mentioned in \cite{Gu-COR24} that it takes too long to compute all feasible matches (hyperedges) in real-time for practical usage. In \cite{Gu-COR24}, a heuristic to reduce the number of hyperedges to be computed is used, and they call it reduction configuration (\textit{Config} for short).
%We need to modify the Config heuristic slightly so that it can apply to our optimization problems.
%A feasible match consists of exactly one driver and one rider is called a \textit{base match}.
%For every driver $\eta_i$ (or rider $r_i$), let $bm(i)$ be the number of base matches $\eta_i$ (or $r_i$) belongs to, after all feasible matches have been computed.
%A Config $(x, y, z)$ means that $bm'(i) \geq \min\{bm(i),x\}$ (where $bm'(i)$ is the number of base matches after reduction), each driver $\eta_i$ can have at most $y$ non-base matches, and each rider $r_i$ belongs to at most $z$ non-base matches of the same driver.
%Our clustering algorithm can reduce the number of hyperedges to be computed in each cluster, and we compare our clustering algorithm with the Config heuristic.

\subsection{Experiment results} \label{experiment-results}
The sets of generated riders and drivers in each interval are the same for all algorithms.
In this section, the ILP~\eqref{formulation-minimization}-\eqref{constraint-binary} is labeled as \textit{ExactMinDist} or \textit{ExactMinNum} (depending on the problem), and we just use \textit{Exact} if it is clear from the context.
A feasible solution means that every rider of $R$ must be \textit{served}, and a driver $\eta_i$ is called an \emph{assigned driver} if there is a feasible match $(\eta_i, R_i)$ in the solution.
The main performance measures of all the algorithms for both problems include incurred travel distance of assigned drivers, the number of assigned drivers, the travel time (duration) of the riders, and computational time.

\subsubsection{MTRS-minDist results}
The overall results are shown in Table~\ref{table-minDist-result}.
Exact labeled with `240' and `720' means that the Exact ILP is allowed to run for 240 and 720 seconds, respectively, to find a solution.
The total incurred travel distance of assigned drivers is the sum of incurred travel distance of all assigned drivers in all intervals.
The average incurred travel distance per assigned driver is calculated as the total incurred travel distance of assigned drivers divided by the total number of assigned drivers.
It is clear that Exact-720 performs the best in terms of incurred travel distance.
Both Exact-240 and Exact-720 have way more assigned personal drivers than that of GreedyMinDist.
Due to the definition of incurred travel distance, it is better to assign riders to personal drivers in general, and this explains the performance difference between Greedy and Exact.
On the other hand, the travel time saved for the riders are very similar between GreedyMinDist, Exact-240 and Exact-720.
The total time saved of all riders is the sum of time saved for all served riders in all intervals, which is around $140800 \pm 200$ minutes depending on the algorithm, as shown in Table~\ref{table-minDist-result}.
The average time saved per rider is calculated as the total time saved of all riders divided by the total number of riders generated, which is about 18.3 minutes.
On average, each served rider saves about 38.6\% of travel time (with acceptance threshold 0.7).
\begin{table}[hb!]
%\footnotesize
\centering
\begin{tabular}{ l | c | c | c }
Total public transit duration & \multicolumn{3}{ l }{365055 minutes} \\
Avg public transit duration per rider & \multicolumn{3}{ l }{47.41 minutes} \\
Total number of drivers generated & \multicolumn{3}{l}{2571 personal and 7700 designated}\\
Total number of riders generated & \multicolumn{3}{l}{7700}\\
\hline
									  & GreedyMinDist & Exact-240 & Exact-720  \\ \hline
Total incurred travel dist of assigned drivers & 36788  & 29400   & 26869 \\
Avg incurred travel dist per assigned driver   & 11.69    & 9.34  & 8.72     \\
Total number of assigned drivers 				& 3220  & 3148  & 3082 	 \\
Avg transit time saved per rider        		& 18.28   & 18.29    & 18.32     \\
%Avg total running time  		  		& 17.72   & 24    & 34.75    \\
%\hline
%With Clustering & GreedyMinDist   & \multicolumn{2}{ c }{Exact (no time limit)}  \\ \hline
%Total incurred travel dist of assigned drivers & 49172  & \multicolumn{2}{ c }{35710} \\
%Avg incurred travel dist per assigned driver   & 13.98   & \multicolumn{2}{ c }{10.64}     \\
%Total time saved of all riders        & 140209  & \multicolumn{2}{ c }{141562}    \\
%Avg time saved per rider        		& 18.21   & \multicolumn{2}{ c }{18.38}     \\
%Avg total running time			       & 10.29   & \multicolumn{2}{ c }{10.61}     \\
\end{tabular}
\caption{Overall solution comparison between all algorithms for all time intervals. Distance is in kilometers and time is in minute.}
\label{table-minDist-result}
\end{table}

Next, we look at the performances of the algorithms using clustering.
Because the running time of the Exact algorithm using clustering is about 30 seconds in the worst case, no time limit was applied to Exact at all.
For clarity, algorithms using clustering are labeled with `C' in front of them.
The results shown in Table~\ref{table-minDist-clustering-result}.
\begin{table}[ht!]
%\footnotesize
\centering
\begin{tabular}{ l | c | c }
   	\hline
		& C-GreedyMinDist   & C-Exact   \\ \hline
Total incurred travel distance of assigned drivers & 49172  & 35710    \\
Avg incurred travel distance per assigned driver   & 13.98    & 10.64    \\
Total number of assigned drivers 				& 3518    & 3355   	    \\
Avg transit time saved per rider        		 & 18.21   & 18.38        \\ \hline
   \end{tabular}
\caption{Results of using clustering. Distance is in kilometers and time is in minute.}
\label{table-minDist-clustering-result}
\end{table}
The number of assigned drivers increases by about 9.25\% for C-GreedyMinDist, compared to GreedyMinDist.
The number of assigned drivers increases by about 6.58\% and 8.86\% for C-Exact, compared Exact-240 and Exact-720, respectively.
The increased in the number of assigned drivers also increases the total incurred travel distance of assigned drivers by about 33.66\%, 21.46\% and 32.9\% for GreedyMinDist, Exact-240 and Exact-720, respectively.
The average incurred travel distance per assigned driver is increased by about 2 km or less for each corresponding algorithm.
The increased in travel distance is more spread out as more drivers are assigned riders.
On the other hand, the time saved for riders is about the same, compared to non-clustering.
Next, we examine whether performance/running time trade-off is meaningful.

Table~\ref{table-minDist-runningTime} shows the computational time for each algorithm discussed above with and without clustering.
For each time interval, we measure (1) the time to compute feasible matches, (2) the time to compute solutions after feasible matches are computed, and (3) the total time to compute feasible matches and solutions.
The table reports the average running times of (1), (2) and (3) per interval (total running time of all 8 intervals divided by 8).
\begin{table}[!ht]
  \centering
  \begin{tabular}{ l | P{3.5cm} | P{2.5cm} | P{2.7cm} }
Without clustering  & Avg computational time of feasible matches & Avg algorithm running time & Total avg running time \\ \hline
GreedyMinDist    & 1062 & 0.742    & 1063	 \\
Exact-240  		 & 1062  & 378.3     & 1440	 \\
Exact-720  		 & 1062  & 1023      & 2085	 \\ \hline
With clustering  & Avg computation time of feasible matches & Avg algorithm running time & Total avg running time \\ \hline
C-GreedyMinDist	& 616.9 & 0.399     & 617.4   \\
C-Exact   		& 616.9 & 19.34     & 636.3   \\
  \end{tabular}
\caption{Running time comparison between all algorithms with and without clustering. Time unit is measured in second.}
\label{table-minDist-runningTime}
\end{table}
For C-Exact, its average running time per interval is only 19 seconds, whereas Exact-240 and Exact-720 are 378 and 1023 seconds, respectively. On average, it takes almost 20 times and 54 times longer than C-Exact for Exact-240 and Exact-720 to run to completion, respectively.
The total average running time gives a better overview.
For GreedyMinDist, Exact-240 and Exact-720, they run to completion in about 17.7 minutes, 24 minutes and 35.75 minutes on average, respectively, whereas C-Exact only needs 10.6 minutes.
C-Exact performs better than GreedyMinDist in both the running time and travel distance.
When comparing C-Exact with Exact-240, C-Exact is about 21.5\% worse than Exact-240 in travel distance, but it reduces the running time by about 56\%. This gaps further increases for C-Exact and Exact-720 (33\% decrease in performance and 69.5\% reduction in running time).

In other words, the average travel time per rider is reduced by about 38.6\% while the average incurred travel distance per driver is less than 9.5 km (for Exact).
When clustering is used, the average travel time per rider is reduced by about 38.7\% while the average incurred travel distance per driver is 10.6 km (for Exact) with 56\% and 70\% running time reduction for Exact-240 and Exact-720, respectively.

\subsubsection{MTRS-minNum results}
As mentioned in the Preliminaries (Section~\ref{sec-preliminary}), the goal of MTRS-minNum is to minimize the number of drives needed to serve all riders, with an emphasis in assigning personal drivers first.
We first compute an assignment $\Pi_{\Gamma}$ that assigns a maximal set $R' \subseteq R$ to drivers of $\Gamma$. Two algorithms described in~\cite{Gu-COR24}: exact ILP (labeled as \textit{ILP-P} here) and \textit{ImpGreedy} are used to compute assignment $\Pi_{\Gamma}$.
Then, we use each of proposed algorithms, LS, GreedyMinNum (\textit{Greedy} for short), and Exact to compute an assignment $\Pi_{\Delta}$ such that all of $R\setminus R'$ are assigned to drivers of $\Delta$.
The results are shown in Table~\ref{table-minNum-result}.
The instance of each time interval is the same for both MTRS-minNum and MTRS-minDist.
The solution $\Pi_{\Gamma}$ computed by ILP-P is about 4.81\% better than that of ImpGreedy, in the number of assigned personal drivers.
Greedy and LS assigns about 12.2\% and 23.8\% more designated drivers than that of Exact, respectively, under ILP-P.
Greedy and LS assigns about 13.7\% and 23.6\% more designated drivers than that of Exact, respectively, under ImpGreedy.
These suggest the relative performance of these three algorithms are somewhat stable.
The time saved for riders are about the same for MTRS-minNum and MTRS-minDist, which is around 140,200 to 141,000 minutes.
The average time saved per rider is about 18.2-18.3 minutes, which equates to about 38.5\% of travel time reduction.
\begin{table}[!t]
\small
\centering
\begin{tabular}{ l | c | c | c | c | c | c }
		& \multicolumn{3}{ c |}{ILP-P} & \multicolumn{3}{ c }{ImpGreedy}  \\ 
	& LS & Greedy & Exact & LS & Greedy & Exact \\ \hline		
Avg incurred travel dist per assigned driver   & 18.73  & 13.23 & 13.03  & 21.92  & 14.16 & 13.93 \\
Total \# of assigned drivers 				& 3324  & 3253  & 3179  		& 3438   & 3359  & 3250  \\
Total \# of assigned personal drivers 		& 2570  & 2570  & 2570  		& 2452	 & 2452   & 2452  \\
Total \# of assigned designated drivers 	& 754   & 683   & 609   		& 986    & 907    & 798   \\
Avg transit time saved per rider        	& 18.33  & 18.3  & 18.26    	& 18.28  & 18.23  & 18.21  \\
\end{tabular}
\caption{Overall solution comparison between all algorithms for all time intervals. Distance is in kilometers and time is in minute.}
\label{table-minNum-result}
\end{table}

Next, we look at the performances of the algorithms using clustering.
The results shown in Table~\ref{table-minNum-clustering-result}.
\begin{table}[ht!]
\small
\centering
\begin{tabular}{ l | c | c | c | c | c | c }
		& \multicolumn{3}{ c |}{ILP-P} & \multicolumn{3}{ c }{ImpGreedy}  \\ 
	& LS & Greedy & Exact & LS & Greedy & Exact \\ \hline		
Avg incurred travel dist per assigned driver   & 16.28  & 16.07 & 15.95  & 16.72  & 16.61 & 16.45 \\
Total \# of assigned drivers 				& 3885  & 3823  & 3732  		& 3810   & 3756  & 3657  \\
Total \# of assigned personal drivers 		& 2466  & 2466  & 2466  		& 2270	 & 2270   & 2270  \\
Total \# of assigned designated drivers 	& 1419   & 1357   & 1266   	& 1540   & 1486   & 1387   \\
Avg transit time saved per rider        	& 18.55   & 18.47    & 18.48   		& 18.49  & 18.47  & 18.40  \\
\end{tabular}
\caption{Results of using clustering. Distance is in kilometers and time is in minute.}
\label{table-minNum-clustering-result}
\end{table}
While using clustering, the number of assigned personal drivers decreases about 4\% for ILP-P and 7.4\% for ImpGreedy, comparing to the corresponding algorithms without using clustering.
This suggests clustering is highly effective for the emphasis of MTRS-minNum.
Having said that, the total number of assigned drivers increase about 17\% for ILP-P and 10-12\% for ImpGreedy, which are within acceptable range.
The increases in total incurred travel distance of assigned drivers are about 43\% for Exact and Greedy (under ILP-P) and 32\% for Exact and Greedy (under ImpGreedy).
These increases were expected, but the magnitudes were higher than expected; however, the emphasis of MTRS-minNum is not incurred travel distance.
The total and average time saved for riders are actually the highest of all results.

Lastly, we look at the running times of the above algorithms.
Table~\ref{table-minNum-ILP-runningTime} and Table~\ref{table-minNum-ImpGreedy-runningTime} show the computational time for each algorithm with and without clustering, where algorithm using clustering is labeled with ``C'' in front of them.
First, we compare the running times of algorithms without using clustering.
The total average running times of Algorithms LS, Greedy and Exact under ILP-P take slightly longer than that when under ImpGreedy (1.5\% to 4\% higher running time).
The performance under ILP-P is about 4.8\% better than that when under ImpGreedy. The trade-off between the running time and performance of ILP-P and ImpGreedy is not significant.
\begin{table}[!th]
\small
\centering
\begin{tabular}{ l | P{3.8cm} | P{2.5cm} | P{2.7cm} }
Without clustering  & Avg computational time of feasible matches & Avg algorithm running time & Total avg running time \\ \hline
ILP-P    		 & N/A  & 16.81   & N/A 	  \\
LS  		     & 857.3  & 3.269   & 877.3	 \\
Greedy  		 & 857.3  & 0.351   & 874.4	 \\
Exact  		     & 857.3  & 79.56   & 953.6	 \\ \hline
With clustering  & Avg computational time of feasible matches & Avg algorithm running time & Total avg running time \\ \hline
C-ILP-P	        & N/A    & 1.548   &  N/A   \\
C-LS   		    & 294.2  & 0.300   &  296.0 \\
C-Greedy	    & 294.2  & 0.040   &  295.8 \\
C-Exact   	    & 294.2  & 2.89    &  298.6 \\
\end{tabular}
\caption{Running time comparison between all algorithms with and without clustering, using ILP-P to compute $\Pi_{\Gamma}$ first. Time unit is measured in second.}
\label{table-minNum-ILP-runningTime}
\end{table}

\begin{table}[!th]
\small
\centering
\begin{tabular}{ l | P{3.8cm} | P{2.5cm} | P{2.7cm} }
Without clustering  & Avg computational time of feasible matches & Avg algorithm running time & Total avg running time \\ \hline
ImpGreedy   & N/A    & 0.0299 &  N/A   \\
LS          & 840.6  & 3.589  &  844.3 \\
Greedy      & 840.6  & 0.242  &  840.9 \\
Exact  	    & 840.6  & 99.46  &  940.1 \\ \hline
With clustering  & Avg computational time of feasible matches & Avg algorithm running time & Total avg running time \\ \hline
C-ImpGreedy & N/A     & 0.0065 &  N/A   \\
C-LS        & 323.7   & 0.485  &  324.2 \\
C-Greedy    & 323.7   & 0.131  &  323.9 \\
C-Exact     & 323.7   & 2.565  &  326.3 \\	\hline
\end{tabular}
\caption{Running time comparison between all algorithms with and without clustering, using ImpGreedy to compute $\Pi_{\Gamma}$ first. Time unit is measured in second.}
\label{table-minNum-ImpGreedy-runningTime}
\end{table}
The total average running times for algorithms C-LS, C-Greedy and C-Exact are about the same: 296 seconds under ILP-P and 324 seconds under ImpGreedy.
Since C-Exact performances the best, C-Exact should be selected among these three algorithms.
When comparing C-Exact with Exact under ILP-P, the total average running time of C-Exact is 68.7\% less than that of Exact, meanwhile C-Exact assigns 4\% fewer personal drivers than that of Exact. 
The total number of assigned drivers increases about 17\% for C-Exact.
The improvement in the running time makes this trade-off meaningful and beneficial.

In summary, the average travel time per rider is reduced by about 38.5\% while 3179 drivers (99.99\% of all personal drivers and 7.9\% of all designated drivers) are required to serve all riders for Exact under ILP-P.
When clustering is used for Exact under ILP-P, the average travel time per rider is reduced by about 39\% while 3730 drivers (95.9\% of all personal drivers and 16.4\% of all designated drivers) are required to serve all riders with 69\% running time reduction.

\section{Conclusion} \label{sec-conclusion}
The proposed system integrates public transit with ridesharing to enhance FM and LM commutes, prioritizing ridesharing routes to improve convenience, reduce commute times, and minimize traffic impacts (while potentially improving traffic conditions).
We give a hypergraph-based ILP exact algorithm and approximation algorithms, along with a clustering algorithm that significantly reduces the time required to compute the hypergraph. Computational results show that our system and algorithms are effective in improving FM/LM commutes.

It will be interesting to see if the approximation ratios of GreedyMinDist and GreedyMinNum can be improved by adapting some local improvement approach, such as incorporating with Algorithm LS or local search algorithms for set packing problems. The clustering algorithm improves the running time significantly, but it also decreases the solution quality notably. It is worth to improve the clustering algorithm so that it can also lead to solution with quality comparable to non-clustering, or to show that it is NP-hard to do so.

\bibliographystyle{splncs04}
\bibliography{iwoca_arxiv}

\end{document}